\documentclass[preprint,review,11pt]{elsarticle}
\usepackage{amssymb,amsmath}
\usepackage{graphicx}
\usepackage{epstopdf}
\usepackage{epsfig}
\usepackage{color}
\usepackage{lineno}
\usepackage{subcaption}
\usepackage{siunitx}
\usepackage{natbib}
\usepackage{lineno}
\usepackage{lscape}
\usepackage{multirow}

\usepackage[version=3]{mhchem}

\usepackage[figuresright]{rotating}

\usepackage[colorlinks=true,
            linkcolor=red,
            urlcolor=blue,
            citecolor=gray]{hyperref}

\journal{Journal of Aerosol Science}

\begin{document}


\begin{frontmatter}



\title{Background PM10 atmosphere: In the seek of a multifractal characterization using complex networks}

\author[ks,ua]{Thomas Plocoste\corref{cor1}}
\ead{thomas.plocoste@karusphere.com}

\author[cg]{Rafael Carmona-Cabezas}
\ead{f12carcr@uco.es}

\author[cg]{Francisco Jos{\'e} Jiménez-Hornero}
\ead{fjhornero@uco.es}

\author[cg]{Eduardo Guti{\'e}rrez de Rav{\'e}}
\ead{eduardo@uco.es}

\cortext[cor1]{Corresponding author}

\address[ks]{Department of Research in Geoscience, KaruSphère SASU, Abymes 97139, Guadeloupe (F.W.I.), France}

\address[ua]{Univ Antilles, LaRGE Laboratoire de Recherche en Géosciences et Energies (EA 4935), F-97100 Pointe-à-Pitre, France}

\address[cg]{Complex Geometry, Patterns and Scaling in Natural and Human Phenomena (GEPENA) Research Group, University of Cordoba, Gregor Mendel Building (3rd ﬂoor), Campus Rabanales, 14071, Cordoba, Spain}


\begin{abstract}

In the literature, several epidemiological studies have already associated respiratory and cardiovascular diseases to acute exposure of mineral dust. However, frail people are also sensitive to chronic exposure to particulate matter with an aerodynamic diameter 10 $\mu$m or less ($PM10$). Consequently, it is crucial to better understand $PM10$ fluctuations at all scales. This study investigates $PM10$ background atmosphere in the Caribbean area according to African dust seasonality with complex network framework. For that purpose, the regular Visibility Graph (VG) and the new Upside-Down Visibility Graph (UDVG) are used for a multifractal analysis. Firstly, concentration vs degree (v-k) plots highlighted that high degree values (hubs behavior) are related to the highest $PM10$ concentrations in VG while hubs is associated to the lowest concentrations in UDVG, i.e. probably the background atmosphere. Then, the degree distribution analysis showed that VG and UDVG difference is reduced for high dust season contrary to the low one. As regards the multifractal analysis, the multifractal degree is higher for the low season in VG while it is higher for the high season in UDVG. The degree distribution behavior and the opposite trend in multifractal degree for UDVG are due to the increase of $PM10$ background atmosphere during the high season, i.e. from May to September. To sum up, UDGV is an efficient tool to perform noise fluctuations analysis in environmental time series where low concentrations play an important role as well.

\end{abstract}

\end{frontmatter}
%

\section{Introduction}	
\label{intro}

	Aeolian processes in North Africa annually transfer important amounts of mineral dust westwards to the Atlantic and Caribbean sea \citep{prospero1981, petit2005, moreno2006, van2016}. Over the ocean, mineral dust transport is made in a Saharan Air Layer (SAL) bounded by temperature inversions and defined by typical vertical gradients of potential temperature and water vapour mixing \citep{prospero1972, adams2012}. Even if dust occurrence is systematically related to a SAL, it is important to underline that a SAL is not always dusty \citep{petit2005}. For the atmosphere to be loaded with soil dust particles, several processes are required in Africa \citep{mahowald2014}: i) strong wind, ii) dry soil, iii) sparse vegetation and iiii) saltating particles.   
	
	Once deposited to the surface by dry or wet deposition \citep{schepanski2018}, dust particles provide micro nutrients to the ocean \citep{martin1991, jickells2005} or to land ecosystems \citep{painter2007, okin2008}. Conversely, mineral dust is also known to have many harmful effects on human health. Indeed, particulate matter with an aerodynamic diameter 10 $\mu$m or less (PM10) are frequently associated to respiratory and cardiovascular diseases \citep{gurung2017, zhang2017, momtazan2019, feng2019}. In the Caribbean area, the health impact of dust outbreaks is frequently related to acute exposure \citep{cadelis2013, cadelis2014}. However, pregnant women, children and the elderly are also sensitive to chronic exposure. Recently, an epidemiological study assessed the impact of dust outbreaks on severe small for gestational-age births in Guadeloupe \citep{viel2020}. The results showed that Saharan dust seems to influence weight but not length or head circumference at birth. Consequently, it is fundamental to understand $PM10$ fluctuations at all scales. Indeed, past chronic exposure studies are usually focused on atmospheric pollutants accumulation effects \citep{woodruff1997, ling2009, scheers2015} while some works pointed out that exposure to short-term fluctuations of air pollution can increase health risks \citep{schwartz1995, maleki2016}.
	
	To perform a profound analysis of $PM10$ time series, multifractal frame is usually used \citep{ho2004, liu2015a, gao2016, dong2017, plocoste2017, plocoste2020d}, to mention a few. In all these studies, large fluctuations of $PM10$ values are frequently taken into account. To our knowledge, no study has yet assessed the background atmosphere of $PM10$ concentrations in a multifractal way. Here, the aim of this study was to perform a profound analysis of $PM10$ noise fluctuations in the Caribbean area according to African dust seasonality with complex network framework. To achieve this, the regular Visibility Graph (VG) and the new Upside-Down Visibility Graph (UDVG) are used.
	
	In order to carry out this study, the paper is organized as follows. Section \ref{MatMeth} describes the data and the theoretical framework applied. Section \ref{results} presents the results obtained and discusses them. Lastly, a conclusion and an outlook for future studies are given in Section \ref{conclusion}.

\section{Experimental data and methods}	
\label{MatMeth}

\subsection{Experimental data}
\label{data}

\begin{figure}[h!]
\centering
\includegraphics[scale=1.0]{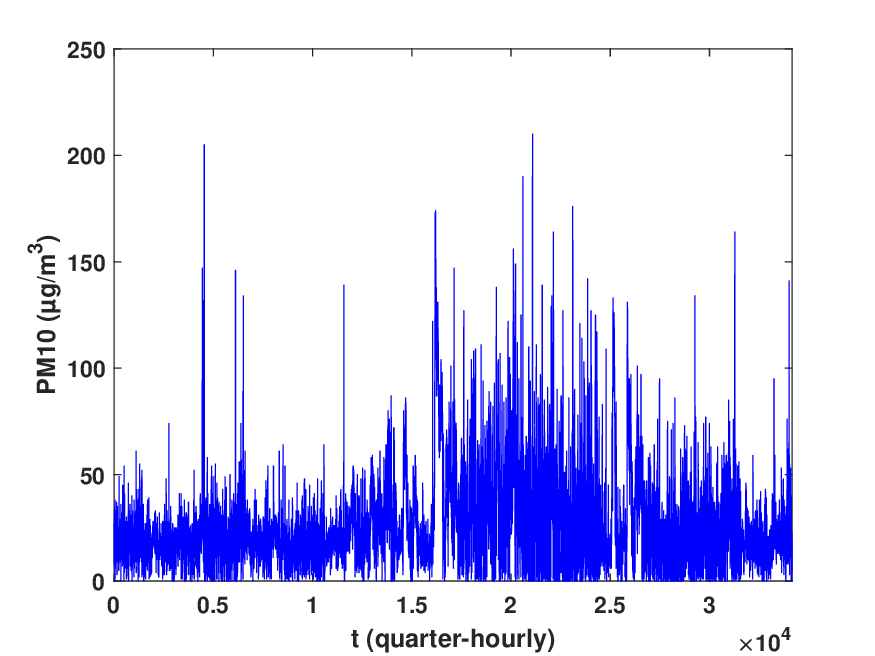}
\caption{\label{signal} $PM10$ time series measured at Pointe-\`a-Pitre in 2009.}
\end{figure}

The time series analyzed here belong to Les Associations Agr\'e\'ees de Surveillance de Qualit\'e de l'air, a national organization that overseas air quality in each of the French administrative regions. In Guadeloupe archipelago ($16.25^\circ$N $-61.58^\circ$W, $\sim$1800 $km^2$), the air quality network is managed by Gwad'Air agency (http://www.gwadair.fr/) \citep{plocoste2019a}. Located at the center of the island, the three air quality stations are close to each other, i.e. less than 10 km for the maximum distance \citep{plocoste2018}. $PM10$ measurements are made using the Thermo Scientific Tapered Element Oscillating Microbalance (TEOM) models 1400ab and 1400-FDMS \citep{prospero2014}. Measurements are made continuously and stored as 15 min averages. Here, we focused on Pointe-\`a-Pitre station ($16.2422^\circ$N $61.5414^\circ$W, urban area) where $PM10$ measurements were performed from 2005 to 2012. Contrary to many studies where hourly data are frequently used \citep{yang2002, alessandro2003, grivas2006, paschalidou2011}, this study deals with quarter-hourly data in order to better investigate $PM10$ fluctuations. To carry out this study, the year 2009 was chosen because it is a classical year (no extreme events) and furthermore the dataset is complete (35040 data points). Figure \ref{signal} depicts the $PM10$ time series where huge fluctuations can be observed, i.e. strong variability.

\subsection{Visibility graphs}
\label{vgmetho}

	Over the past decade, a new technique that transforms time series into graph or network has been developed. Called Visibility Graph (VG) due to its similarities with those used in architecture for space analysis \citep{turner2001}, VG applied to time series was firstly introduced by \cite{lacasa2008}. In the literature, numerous studies have already shown that VG has the benefit of inheriting properties from the original time series \citep{lacasa2008, lacasa2009, lacasa2010, mali2018, carmona2019a}, to mention a few.
	
	Usually, a graph can be described as a set of vertices, points or nodes connected to each other by lines that are usually called edges. In VG frame, the points in the time  series are represented by the nodes. In order to transform a time series into VG, a criterion must be established for linking the nodes and establishing the edges. The concept is the following: two nodes are linked to each other if and only if a line between them can be drawn directly, i.e. without passing below any other point in the signal. Thus, for a time series  $y(t)$, two points ($t_a$, $y_a$) and ($t_b$, $y_b$) will be connected in the graph (have visibility), if any given point ($t_c$, $y_c$) between them ($t_a < t_c < t_b$) meets the following condition \citep{lacasa2008}:

\begin{equation}
y_{c} < y_{a} + (y_{b}-y_{a}) \frac{t_{c}-t_{a}}{t_{b}-t_{a}}
\label{VG}
\end{equation}

	According to VG frame, nodes with highest connectivity (so-called hubs) are frequently associated to highest values in the original time series \citep{carmona2019b}. In order to analyze the large fluctuations in a time series, VG is a very robust tool. However, to study small fluctuations, this method has some drawbacks. This is the reason why this approach is not suitable for studying the behavior of the background atmosphere in air pollution field.
	
	To remedy this, a new version of VG method was recently presented by \cite{soni2019} in order to retrieve more information from a time series. In this new frame, the concept of a signed complex network is integrated. Here, the main concept through this new approach is that some of the edges will have a positive sign, while some other will be negative. Thus, the classical VG firstly presented corresponds to the positive edges of this signed graph while the negative connections are made also from the classical VG but computed this time over the “upside-down” time series. In more concrete terms, the converted series $-f(t)$ is employed in place of the original series $f(t)$. This new method has already proven reliable in different fields \citep{lacasa2015, sannino2017, carmona2020a}.
	
\begin{figure}[h!]
\centering
\includegraphics[scale=0.50]{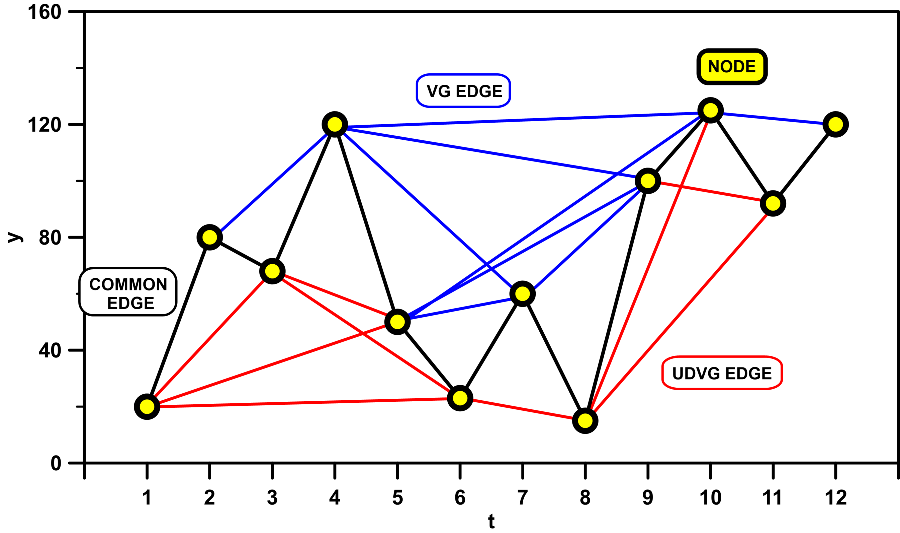}
\caption{\label{visibility} Illustration of computation for the regular VG (blue lines) and the UDVG (red lines) to a sample time series and resulting graphs. Black lines show the common edges.}
\end{figure}

	The aim of this study is to investigate the background atmosphere behavior related to African dust haze in the Caribbean area. In order to perform a profound analysis of this entity, the positive and negative parts need to be obtained separately. Therefore, for the rest of the manuscript, the “positive” network will be termed regular VG while the “negative” one will be called Upside-Down VG (UDVG). Figure \ref{visibility} depicts an example with both networks. As it can be observed, the edges of both graphs differ (VG in blue lines and UDVG in red lines) except those linking each node to it closest neighbors (in black lines) in the time series. This is reflected in the adjacency matrix as well. The last one is one of the most usual ways of representing a graph, which consists of a $N\,\times\,N$ binary matrix (being $N$ the number of nodes or points in the time series). The element $a_{ij}$ of the matrix equals 1 when the nodes $i$ and $j$ and zero otherwise. Therefore, the adjacency matrix of an undirected graph is symmetric (as it is the case). As introduced before regarding VG and UDVG matrices, they meet the following criteria \citep{carmona2020a}: $a_{ij}^{VG}\,+\,a_{ij}^{UDVG} \leq 1$; $\forall j \neq i \pm 1$. Consequently, for both matrices, the elements surrounding the main diagonal are identical and the others cannot be $a_{ij} = 1$ at the same time. In this study, the computational cost of VG and UDVG algorithms is respectively 1.7429 and 1.5333 seconds for the entire $PM10$ time series. 
	
\subsection{Multifractal analysis}
\label{methodMulti}

	In early 80s, multifractal analysis was firstly introduced by \cite{mandelbrot1982} in order to study the energy dissipation in the context of the fully developed turbulence with multiplicatives cascades models. Subsequently, this approach was widely used in environmental studies \citep{tessier1994, seuront1996, kravchenko1999, lee2002, jimenez2011, calif2014, baranowski2015, dong2017, carmona2019b, plocoste2020a}, to cite a few. Indeed, due to the possibility of having different densities depending on the region of application, multifractal analysis is regarded as the inherent property of complex and composite systems \citep{mandelbrot1974}. Classically, two methods are frequently used to analyze the multifractal properties of a time series: the generalized fractal dimension $D_q$ \citep{tel1989, block1990, schreiber1991, posadas2001} and the singularity spectrum $f(\alpha)$ \citep{chhabra1989, bacry1993, lyra1998, caniego2005}.
	
\subsubsection{Generalized fractal dimension}
\label{methodGen}	
	
	The generalized fractal or Rényi dimensions $D_q$ was the first approach to investigate multifractal formalism \citep{harte2001}. This method highlights the scaling exponents of the $qth$ moments of the system \citep{feder1988}. Usually, the fixed size algorithms (FSA) is applied to perform a multifractal analysis \citep{halsey1986, mach1995}. However, FSA does not correctly estimate the side corresponding to negatives value of $q$. To overcome this drawback, the sandbox algorithm (SBA) is introduced by \cite{tel1989}. Based on the box-counting algorithm \citep{halsey1986}, SBA is able to reliably computing the fractal dimensions of real data, even for negative moments \citep{tel1989}. Initially developed by \cite{vicsek1990}, this approach was firstly applied in complex networks frame for multifractal analysis by \cite{liu2015b}. Many studies have shown that SBA is the most effective, feasible and accurate algorithm to investigate the multifractal behavior and compute the mass exponent of complex networks \citep{yu2016, mali2018, carmona2019b}.
	
	In SBA procedure, a number of randomly placed boxes are selected for each radius. These boxes are always centred on a node of the network. As a consequence, the entire network is covered with those boxes by choosing a sufficiently high number of them \citep{carmona2019b}. The following equation is used to determine the probability measurement to compute each box ($B$) \citep{carmona2019b}:
		
\begin{equation}
\mu(B) = \frac{M(B)}{M_0}
\label{SBA1}
\end{equation}

Being M(B) the number of counted boxes or points in a given sandbox with radius $r$ and $M_0$, the total amount for the whole fractal object. After computing this value for each box and radii, the classical generalized fractal dimensions can be obtained for the different $q$ values as follows \citep{carmona2019b}: 

\begin{equation}
D_{q} = \frac{1}{q-1}\, \lim_{r\to0} \frac{ln\left<\mu(B)^{(q-1)}\right>}{ln\,r}\,\forall q \neq1
\label{SBA2}
\end{equation}

Where the $<->$ terminology indicates that all the $\mu(B)$ values from the randomly generated boxes are averaged for each radii $r$. For the case $q=1$, one can obtain the following expression \citep{mali2018}:

\begin{equation}
D_{1}= \lim_{r\to0} \frac{<ln\mu(B)>}{ln\,r}
\label{SBA3}
\end{equation}

	In the literature, the protocol for executing the SBA algorithm for complex networks is widely described \citep{liu2015b, yu2016, carmona2019b}. Here, the input parameters are the following: i) the interval used for the radii goes from 1 to 30 ($r \in [1, 30]$) according to the distance matrix between the nodes; ii) the range of moments is between $q = -5$ and $q = +5$ with an increment step of 0.25. From equation \ref{SBA2}, numerous informations can be extracted \citep{carmona2019b}: i) $D_{q=0}$ corresponds to the fractal dimension of the given system or box-counting dimension; ii) $D_{q=1}$ is the so-called information entropy; iii) $D_{q=2}$ describes the correlation dimension. Clasically, multifractality degree can be estimated by $\Delta\,D_q = max\,D_q - min\,D_q$ \citep{yu2016}.

\subsubsection{Singularity spectrum}
\label{methodSing}

	The singularity or multifractal spectrum is another approach to investigate multifractal characteristics of a time series. In many studies, the Legendre transformation from mass exponents $\tau(q)$ have been applied to compute it \citep{muzy1993, olsen1995, schmitt2005, calif2013}. However, the possible inclusion of spurious points and error amplification from the derivative are serious constraints in applying this transformation \citep{chhabra1989, veneziano1995}. Moreover, as $\tau(q) = (1 - q)D_q$, Legendre transform is dependent to Rényi spectrum. To overcome those limitations, a new way to determine the $\alpha$-spectrum directly from the original time series was introduced by \cite{chhabra1989}. To calculate the probabilities of the boxes of radius $r$, this approach is based on the normalized measure $\beta_i(q)$ and $\mu_i$ from the original time series with the following formula \citep{chhabra1989}:     

\begin{equation}
\beta_i(q,r) = [P_i(r)]^q\,/\,\sum_{j}[P_j(r)]^q
\label{sing1}
\end{equation}

with $P_i(r)$ the different fractal measurements for each box of radius $r$, i.e. the number of nodes. Subsequently, from Equation \ref{sing1}, $f(\alpha)$ and $\alpha$ are obtained by using the following formulas \citep{chhabra1989}: 

\begin{equation}
f(q) = \lim_{r\to0} \frac{\sum_{i}\,\beta_i(q,r)\,log[\beta_i(q,r)]}{log\,r}
\label{sing2}
\end{equation}

\begin{equation}
\alpha(q) = \lim_{r\to0} \frac{\sum_{i}\,\beta_i(q,r)\,log[P_i(r)]}{log\,r}
\label{sing3}
\end{equation}

where $\alpha$ is the Lipschitz-Hölder exponent \citep{posadas2001}. In more concrete terms, those elements are determined using the slope of $\sum_{i}\,\beta_i(q,r)\,log[\beta_i(q,r)]$ over $log\,r$ and $\sum_{i}\,\beta_i(q,r)\,log[P_i(r)]$ over $log\,r$ respectively for $f(\alpha)$ and $\alpha(q)$. This slope is defined by means of a linear regression in the same range of radii where the other fractal measures are computed \citep{carmona2019b}. Usually, multifractality degree can be estimated by the width of the spectrum $W = \alpha_{max} - \alpha_{min}$ \citep{mali2018}.

\section{Results and Discussion}
\label{results}

\subsection{Overall analysis}
\label{resultOver}

\subsubsection{Degree distribution}
\label{resultOverDeg}

\begin{figure}[h!]
\centering
\includegraphics[scale=0.55]{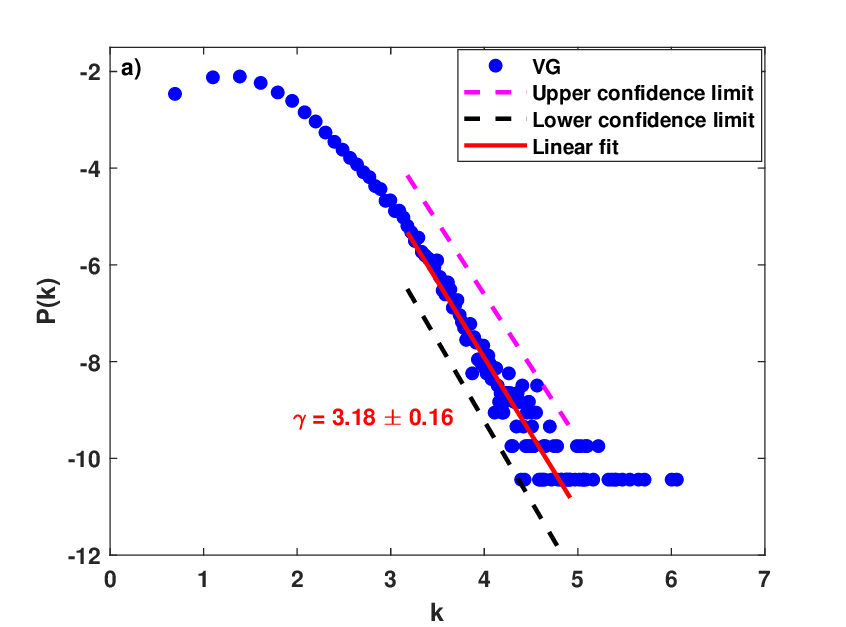}
\includegraphics[scale=0.55]{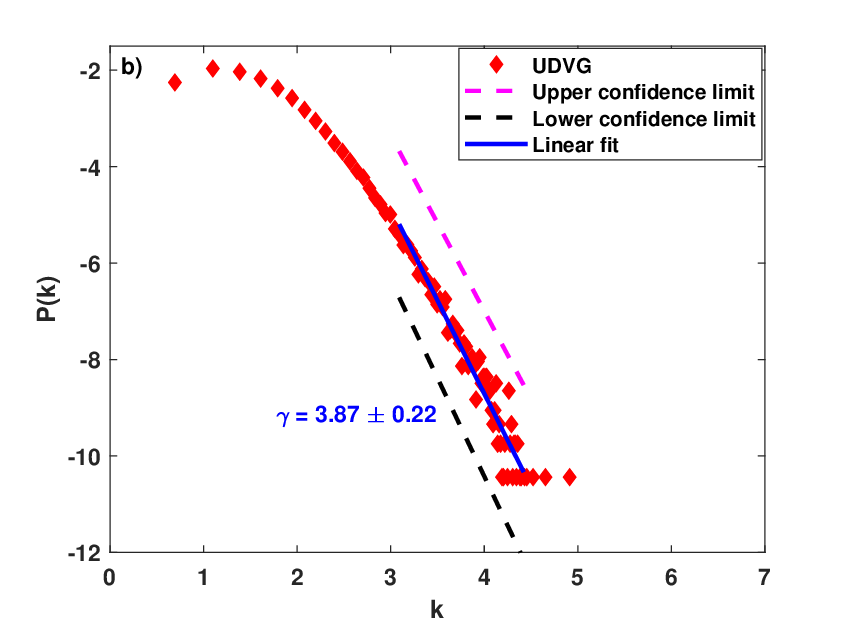}
\includegraphics[scale=0.55]{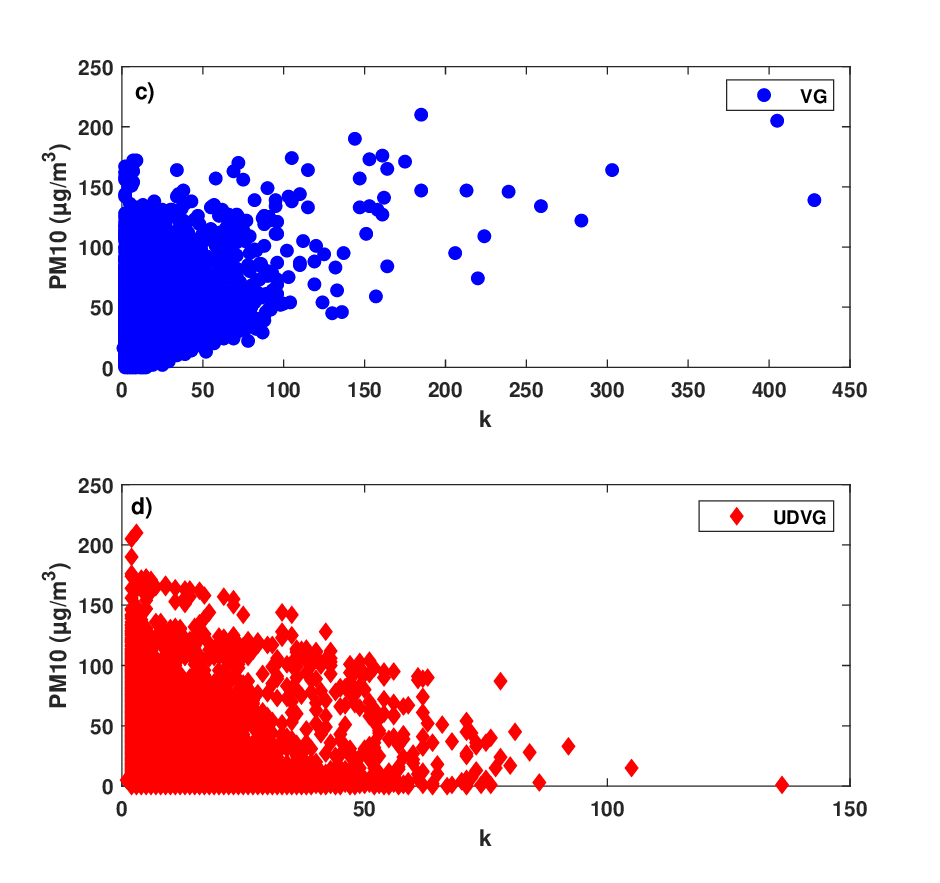}
\caption{\label{degOver} Degree distribution for the overall $PM10$ data in (a) Visibility Graph (VG) and (b) Upside-Down Visibility Graph (UDVG) frames. All tails of degree distribution are fitted by a linear regression with confidence interval at 90\%. (c) and (d) depict the relationship between $PM10$ time series values and their degrees respectively for VG and UDVG methods.}
\end{figure}

	To start this study, the degree distribution $P(k)$ behavior between the regular VG and UDVG has firstly been analyzed by the authors for $PM10$ time series. To achieve this, Figure \ref{degOver} depicts in (a)-(b) the degree distribution for all data in VG and UDVG frames and in (c)-(d) the values of $PM10$ concentrations plotted against the degree respectively for VG and UDVG. 
	
	In Figure \ref{degOver}(a)-(b), one can observe that VG and UDVG degree distributions are almost coincident for low degree values. This results are consistent with concentration vs degree (v-k) plots in Figure \ref{degOver}(c)-(d). Indeed, low degree values are mostly related to intermediate $PM10$ concentrations in both cases.
		
	For high $PM10$ concentrations (hubs behavior), the distributions are different. In both cases, the tail region of the log-log plot of $P(k)$ can be fitted by a power law like $P(k) \propto  k^{-\gamma}$ but the $\gamma$ exponent differs considerably. Thus, the exponents computed from the slopes are estimated for $k \geq 22$ with respectively $3.18 \pm 0.16$ and $3.87 \pm 0.22$ for VG and UDVG. The same behavior between VG and UDVG degree distribution was previously observed by \cite{carmona2020a} for nitrogen dioxide ($NO_2$) in Cadiz, Spain. After computing the skewness for $PM10$ data ($S_{PM10}=2.33 \pm 0.03$), it is important to emphasize that $NO_2$ data exhibit the same skewness value with $S_{NO_2}=2.32 \pm 0.05$. For tropospheric ozone ($O_3$) data, \cite{carmona2020a} found that VG and UDVG degree distributions are closer with skewness value equal to -0.28, i.e. symmetrical distribution. The pollutants nature could explain these skewness values. Indeed, contrary to $O_3$ which is a secondary pollutant, $PM10$ and $NO_2$ measured in Guadeloupe archipelago come mainly from primary sources \citep{plocoste2018}. In other words, hubs behavior seems to be closely linked to skewness value.
	
	In many studies, v-k plot has shown that highest degrees (hubs) are related to the largest values in the regular VG frame \citep{pierini2012, carmona2019b}. Looking at Figure \ref{degOver}(c), one can notice that this trend is confirmed. As expected for UDVG, Figure \ref{degOver}(d) highlights the opposite behavior. In this case, the degree clearly decays as $PM10$ concentration rises and v-k relationship is even clearer and smoother that in VG case. UDVG hubs seem to describe the $PM10$ background atmosphere because this approach take into account fluctuations in low values. This assumption will be discussed later on.

\subsubsection{Multifractal analysis}
\label{resultOverMulti}
	
	In order to estimate the multifractal properties of $PM10$ time series between the regular VG and UDVG, the Rényi and the singularity spectra are computed.\\
	
\begin{figure}[h!]
\centering
\includegraphics[scale=0.85]{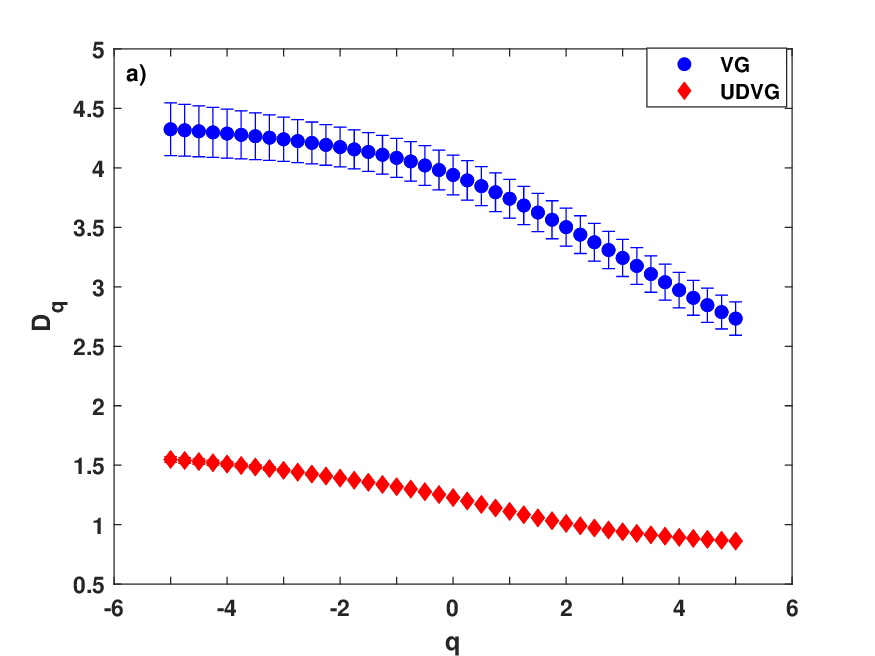}
\includegraphics[scale=0.85]{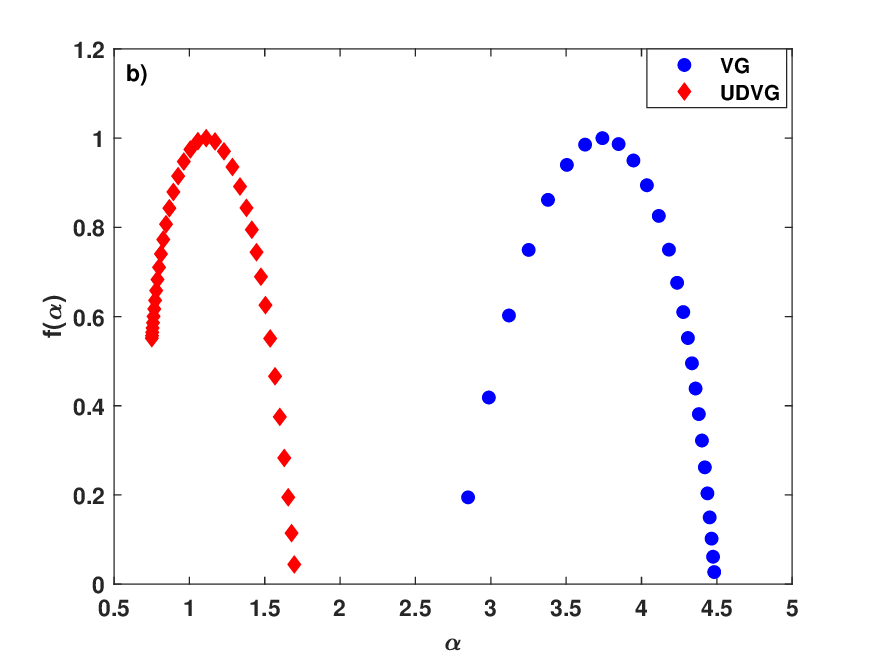}
\caption{\label{multiOver} Illustration of (a) Rényi dimensions and (b) singularity spectrum for the overall in VG and UDVG frames. Standard deviations are illustrated by the whiskers.}
\end{figure}

	For Rényi spectrum, the SBA procedure is firstly applied. Therefore, the elements $\frac{ln\left<\mu(B)^{(q-1)}\right>}{q-1}$ for $q \neq1$ and $<ln\mu(B)>$ for $q = 1$ are used against $ln\,r$. From the partition functions, a linear regression is performed for all $q$ values in order to build Rényi spectrum from Equation \ref{SBA2} and \ref{SBA3}. From $q = -5$ to $q = +5$ with an increment step of 0.25, a linear regression was performed for $0 \leqslant ln\,r \leqslant 2.5$ in VG and $0 \leqslant ln\,r \leqslant 3.5$ in UDVG. This methodology is widely described and illustrated in the literature \citep{liu2015b, yu2016, mali2018, carmona2019b}. Figure \ref{multiOver}(a) shows the Rényi spectrum for both approaches. One can observe the multifractal properties of $PM10$ time series in both cases because $D_0 > D_1 > D_2$ (see Table \ref{VGresult}). However, a clear difference between VG and UDVG behavior is noticed. For UDVG, $D_0$ value is remarkably lower than that of the regular VG. This result is consistent because $D_0$ is related to how the fractal object is covered. We assume that the shape of the lower envelope of the $PM10$ concentration is responsible for this decay in $D_0$ value. In this case, it is more difficult for hubs to “see” nodes that are further away and hence, the degree is lowered. This effect is visible in the v-k plot (see Figure \ref{degOver}(b)), as the VG hubs have much higher degrees.
	
	 Multifractal degree ($\Delta\,D_q$) is higher for VG than UDVG. In other words, fluctuations for high concentrations are more important than fluctuations in low concentrations. These first results show that UDVG frame represents the background atmosphere. Indeed, it seems that the background atmosphere concentration of $PM10$ has a less multifractal behavior, which ends up in a flatter Rényi curve. This is consistent because in insular context, the background atmosphere is mainly composed of marine aerosols and anthropogenic pollution \citep{clergue2015, rastelli2017} which are constant $PM10$ sources through the year. Marine aerosols will be advected by the trade winds which blow continuously during the year \citep{plocoste2014, plocoste2020e} while anthropogenic pollution is produced by the daily human activities \citep{plocoste2018}. Consequently, the background atmosphere might have more regular and stable dynamics through the year. This is confirmed by the fact that standard deviation values (the whiskers) are weak for UDVG.	 
	 The differences $D_{0} - D_{1}$ and $D_{0} - D_{2}$ are as well lower for UDVG. As expected, $D_{0} - D_{1}$ is higher for VG because hubs are related to strong $PM10$ concentrations which are mainly due to dust outbreaks \citep{prospero2014, euphrasie2020, plocoste2020b}, i.e. the upper envelope has a much more irregular and volatile behavior. As regards $D_{0} - D_{2}$, the meaning of this parameter is less clear in this context. Here, we assume that this parameter could correspond to fluctuation degree. However, further studies will be needed to confirm this.\\
	 	 
	 For singularity spectrum, the \cite{chhabra1989} procedure is firstly required. Thus, the elements $\sum\,\beta_i(q,r)\,ln[\beta_i(q,r)]$ and $\sum\,\beta_i(q,r)\,ln[P_i(r)]$ have been used against $ln\,r$ respectively for computing $f(\alpha)$ and $\alpha$. For better comparison, the same range of linear regression as Rényi spectrum has been chosen, i.e. $0 \leqslant ln\,r \leqslant 2.5$ for VG and $0 \leqslant ln\,r \leqslant 3.5$ for UDVG. The steps to build singularity spectrum are widely described in literature \citep{kelty2013, mali2018, carmona2019b}. 
	 
	 Figure \ref{multiOver}(b) shows the singularity spectrum for both approaches. Overall, UDVG spectrum values seem more homogeneous than VG because the fluctuations are less important for low concentrations. Multifractal degree ($W$)values are consistent with Rényi spectrum, i.e. $W_{UDVG} < W_{VG}$. Thus, UDVG is marked by a less multifractal dynamics than VG.
	 
	 According to the multifractal theory, the two sides of the $f(\alpha)$ spectrum are related to different scales in the signal. While the left part (related to $q > 0$) filters out the large fluctuations, the right side ($q < 0$) corresponds to small noise-like variations \citep{mali2018}. Looking at the spectra mentioned before, it is possible to notice the asymmetry of their shapes, again highlighting a difference between the nature of the VG and UDVG, in consonance with what was seen before. Based on the behavior of the hubs (previously discussed in section \ref{resultOverDeg}), the left tail of $f(\alpha)$ is more elongated for VG. This could be expected, as it might be highlighting the multifractal nature of the large fluctuations for the VG and the small noise-like fluctuations for the UDVG, being in this case associated to the background pollutant concentration.\\
	 	 
	 In the following section, multifractal properties of $PM10$ times series are analyzed in VG and UDVG frames according to African dust seasonality.

\begin{landscape}
\begin{table}[!p]
\small
\begin{tabular}{c |c c c c c c c| c c c c c c c|}
\hline
\multicolumn{1}{c}{}& \multicolumn{7}{|c|}{\textbf{VG}} & \multicolumn{7}{c|}{\textbf{UDVG}}\\
\hline
\multicolumn{1}{c}{Period}& \multicolumn{1}{|c}{$D_0$} & \multicolumn{1}{c}{$D_1$}&\multicolumn{1}{c}{$D_2$}& \multicolumn{1}{c}{$D_0 - D_1$} & \multicolumn{1}{c}{$D_0 - D_2$}&\multicolumn{1}{c}{$\Delta\,D_q$}&\multicolumn{1}{c|}{$W$}& \multicolumn{1}{c}{$D_0$} & \multicolumn{1}{c}{$D_1$}&\multicolumn{1}{c}{$D_2$}& \multicolumn{1}{c}{$D_0 - D_1$} & \multicolumn{1}{c}{$D_0 - D_2$}&\multicolumn{1}{c}{$\Delta\,D_q$}&\multicolumn{1}{c|}{$W$} \\
\hline
Overall & 3.940 & 3.741 & 3.502 & 0.200 & 0.439 & 1.592 & 1.640 & 1.227 & 1.111 & 1.010 & 0.116 & 0.217 & 0.686 & 0.945\\
\hline 
Low season & 3.833 & 3.636 & 3.393 & 0.197 & 0.440 & 1.547 & 1.582 & 1.816 & 1.751 & 1.672 & 0.066 & 0.144 & 0.523 & 0.818\\
\hline
High season & 3.180 & 2.997 & 2.794 & 0.183 & 0.386 & 1.311 & 1.458 & 1.274 & 1.183 & 1.102 & 0.092 & 0.173 & 0.686 & 1.037\\
\hline
\end{tabular}
\caption{\label{VGresult} Multifractal parameters from VG and UDGV frame for all data, low dust season (October to April) and high dust season (May to September).}
\end{table}
\end{landscape}

\subsection{Seasonal analysis}
\label{resultSeason}

\subsubsection{Degree distribution}
\label{resultSeasonDeg}

\begin{figure}[h!]
\centering
\includegraphics[scale=0.45]{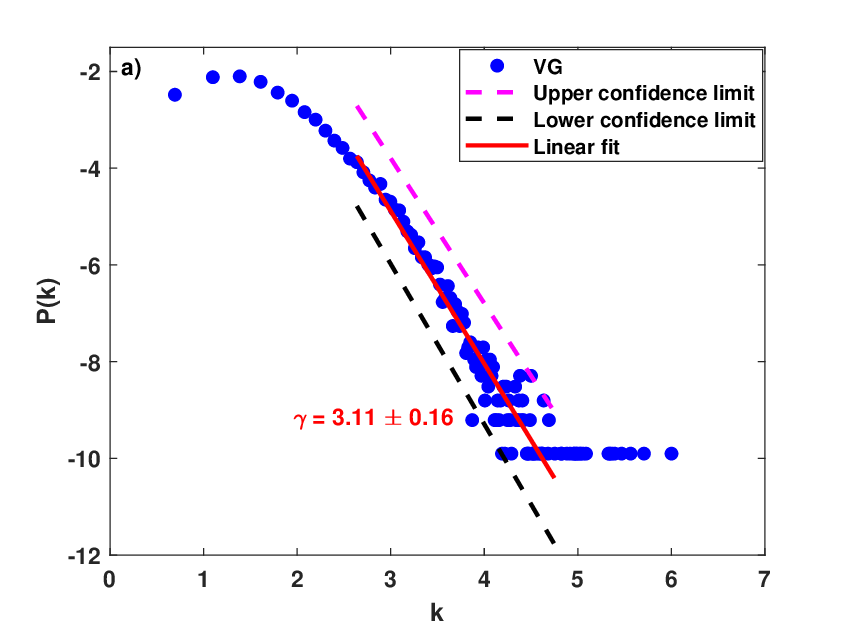}
\includegraphics[scale=0.45]{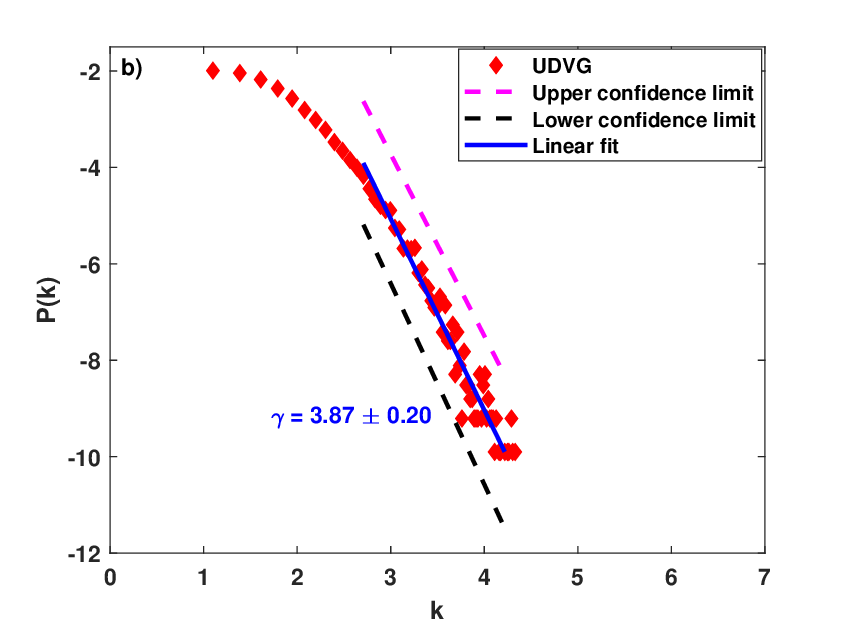}
\includegraphics[scale=0.45]{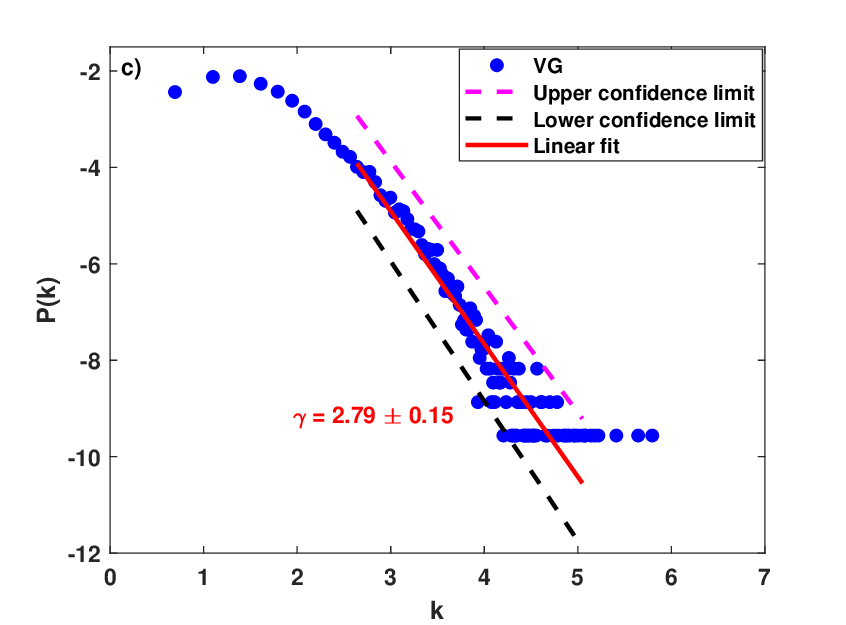}
\includegraphics[scale=0.45]{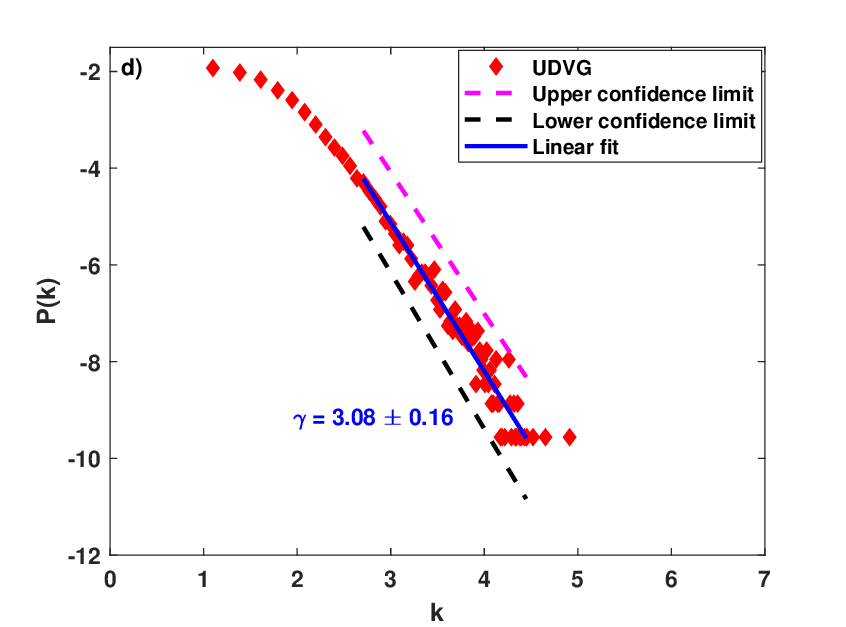}
\caption{\label{degSeas} Degree distribution for the low dust season (October to April) in (a) Visibility Graph (VG) and (b) Upside-Down Visibility Graph (UDVG) frames; and for the high dust season (May to September) in (c) Visibility Graph (VG) and (d) Upside-Down Visibility Graph (UDVG) methods. All tails of degree distribution are fitted by a linear regression with confidence interval at 90\%.}
\end{figure}

	Here, the aim is to investigate degree distribution $P(k)$ behavior between the regular VG and UDVG according to African dust seasonality, i.e. the low dust season (October to April) and the high dust season (May to September) \citep{prospero2014, plocoste2020a}.
	 Figure \ref{degSeas}(a)-(d) illustrates the achieved results. As expected, the low degrees of VG and UDVG coincide for both low and high dust seasons since they are related mostly to intermediate concentrations. On the other hand, there is a difference for high degrees of VG and UDVG. The tail region of the log-log plot of $P(k)$ can be fitted by a power law $P(k) \propto  k^{-\gamma}$ and the gamma exponents computed from the slopes are estimated for $k \geq 13$ with respectively $3.11 \pm 0.16 $/$3.97 \pm 0.20$ for VG/UDVG in the low season and $2.79 \pm 0.15$/$3.08 \pm 0.16$ for VG/UDVG in the high season. The skewness is respectively equal to $2.52 \pm 0.04$ and $1.71 \pm 0.04$ from October to April and May to September. During the high dust season, one can notice that the skewness and the gap between VG-UDVG gamma exponents values are lower. This may be due to the fact that dust outbreaks are more frequent from May to September, i.e. more days with high $PM10$ concentrations \citep{plocoste2020c}. Indeed, during summer months, there is an average of $\sim$6 dust outbreak days in a month \citep{huang2010}. The arrival of a dust outbreak is preceded by the passage of African Easterly Waves (AEWs) \citep{prospero1981} whose frequency is 3-5 days at scale of 2000-3000 $km$ \citep{burpee1972, karyampudi1988, prospero2003}. In the literature, AEWs is also called “African” disturbances because of their sub-Saharan origin \citep{carlson1969}. Dust plume is confined between two consecutive AEWs and the SAL top and base inversions \citep{karyampudi1999}. Due to the high amount of dust in the outbreaks, the atmosphere is often quite turbid in the Caribbean area \citep{prospero1981, karyampudi1999}. In other words, there is still a residual quantity of dust in the atmosphere due to the continuous alternation between AEWs and dust outbreaks. A $PM10$ statistical analysis made by \cite{plocoste2020b} in Guadeloupe archipelago over one decade confirmed this trend. This study highlighted that the maximum probability value from the Probability Density Function (PDF peak), goes from 16.9 $\mu g/m^3$ to 19.6 $\mu g/m^3$ between the low season and the high dust season. In summer, whatever the source or process, the dust is carried into the Caribbean. To sum up, the large pulses of dust are often associated with easterly waves. There is a general background of dust in summer months that is linked to various processes that generate and transport dust to the Atlantic. Consequently, the background $PM10$ atmosphere is higher during summer and UDVG becomes more similar to what VG is describing.

\subsubsection{Multifractal analysis}
\label{resultSeaMulti}

	To perform the multifractal analysis with VG and UDVG frames according to African dust seasonality, the same procedures and linear regression ranges used in the section \ref {resultOverMulti} were applied for the two multifractal approaches.\\
	
\begin{figure}[h!]
\centering
\includegraphics[scale=0.75]{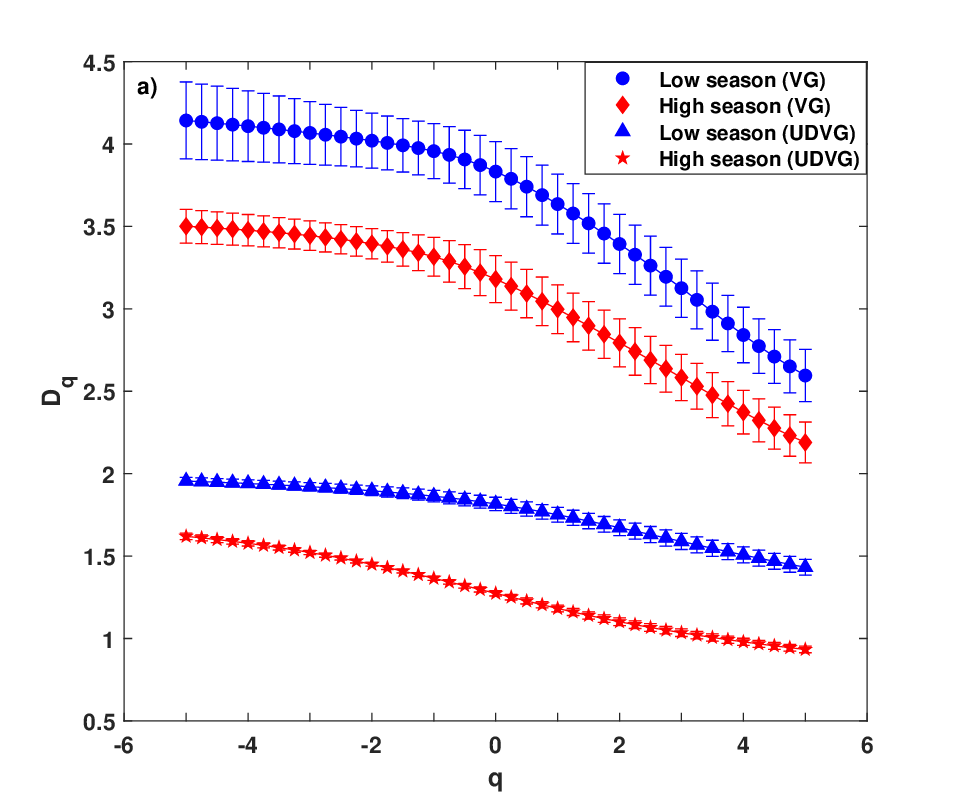}
\includegraphics[scale=0.75]{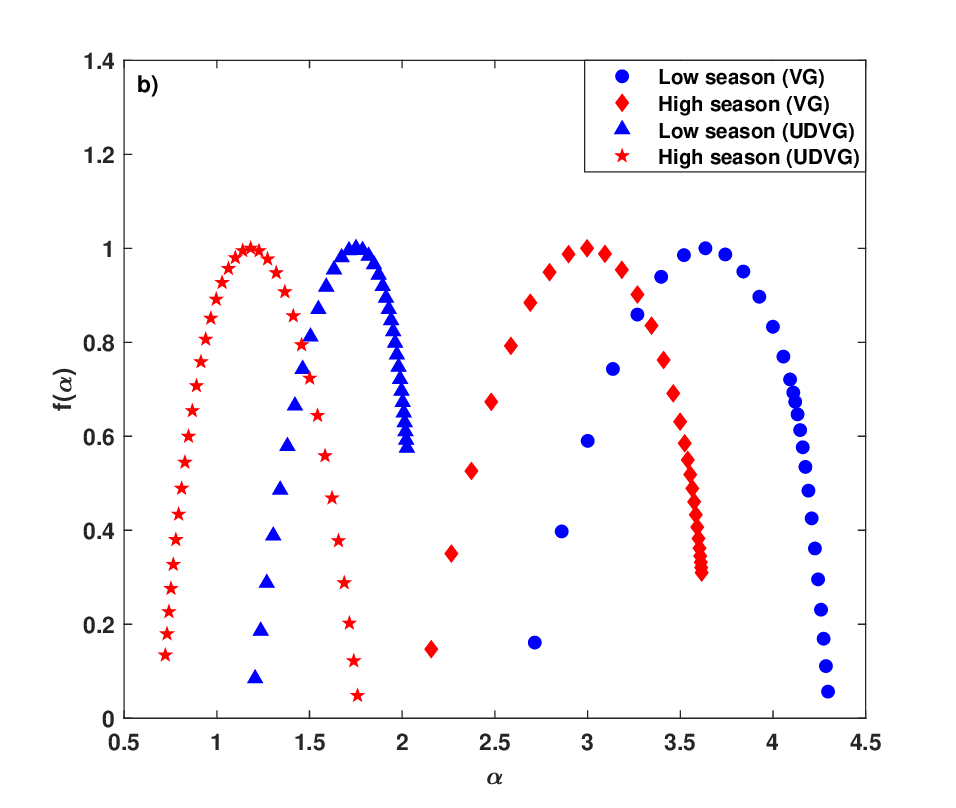}
\caption{\label{MultiSeas} Illustration of (a) Rényi dimensions and (b) singularity spectrum for low dust season (October to April) and high dust season (May to September) in VG and UDVG frames. Standard deviations are illustrated by the whiskers.}
\end{figure}

	Figure \ref{MultiSeas}(a) depicts the Rényi spectra obtained for both seasons. As expected, Rényi spectra in VG have higher values than UDVG. Whatever the season, flatter curves and weaker $D_0$ values are observed for UDVG frame (see Table \ref{VGresult}). As the overall case, multifractal degree ($\Delta\,D_q$) is higher for VG than UDVG.
		
	For VG, the low dust season curve exhibits higher values and higher multifractality degree ($\Delta\,D_q$). Between both seasons, the high dust season is the most uniform period because its exhibits the lowest $D_{0} - D_{1}$ values. According to \cite{prospero2003}, in summer, satellite images show dust outbreaks that emerge from the west coast of Africa in pulses every 3 to 5 days, following behind AEWs. Thereafter, 5 to 7 days later, dust cloud reaches the Caribbean basin \citep{velasco2018}. From October to April, dust outbreaks are more sporadic. Only a particular circulation of air masses in spring \citep{jury2017} or extreme events such as a volcanic eruption \citep{plocoste2019b} can bring dust haze. On the other hand, recurrence degree is higher during the low dust season because it exhibits the maximum value of $D_{0} - D_{2}$. This is explained by the fact that between October and April, the $PM10$ concentrations are mainly linked to anthropogenic activity and marine aerosols which composed the background atmosphere \citep{clergue2015, rastelli2017}. Overall, the same trend was observed in our previous study but there are some differences \citep{plocoste2020d}. This may be attributed to the difference in $PM10$ data resolution as the whiskers seem significantly smaller, i.e. a more accurate description than in our previous study where daily data is used \citep{plocoste2020d}.
	
	As regards UDVG, the low dust season curve still exhibits higher values but multifractality degree ($\Delta\,D_q$) is now higher for the high dust season. Between UDVG and VG there is a change in multifractality degree seasonal trend, which is flatter for the low season and gains multifractal degree in the high dust season due to the general background of dust in summer months \citep{prospero1981, karyampudi1999}. The quantities $D_{0} - D_{1}$ and $D_{0} - D_{2}$ also show an inverted comparison, with respect to the VG case. Due to the alternation in continuously between AEWs and dust outbreaks, background $PM10$ atmosphere concentrations fluctuate more.\\
	
	Figure \ref{MultiSeas}(b) illustrates the singularity spectra obtained for both seasons. In Table \ref{VGresult}, one can underline that $W$ values are in agreement with $\Delta\,D_q$ values for each case. 
	
	For VG, the right tail of $f(\alpha)$ spectrum is more extended for the low season while the left tail values are more compact for the high season. Consequently, the small noise-like fluctuations are more probable for the low season while large fluctuations are more likely in the high season. These results are consistent with our previous findings \citep{plocoste2020d}.

	Regarding UDVG, one can observe that the right tail of $f(\alpha)$ spectrum is more extended for the high season and more homogeneous for the low season. This indicates the persistence behavior of $PM10$ concentrations in the background atmosphere for the low season with marine aerosols and anthropogenic pollution. On the other hand, the left tail of $f(\alpha)$ for the high season is more homogeneous. Here, the increase of fluctuations in background atmosphere values is highlighted from May to September. All these results show the existence of a general background of dust in the high season.

\section{Conclusion}
\label{conclusion}

	In the literature, the health impact of dust outbreaks is frequently related to acute exposure. However, frail people like pregnant women, children and the elderly are also sensitive to chronic exposure. The aim of this paper was to perform a profound analysis of particulate matter ($PM10$) background atmosphere in the Caribbean area according to African dust seasonality with complex network framework. To achieve this, the regular Visibility Graph (VG) and the new Upside-Down Visibility Graph (UDVG) are used.
	
	Firstly, the degree distribution analysis between VG and UDVG methods is performed for the whole year. For low degree values, VG and UDVG degree distributions are almost coincident while they are different for high degree values (hubs behavior). Consequently, the gamma ($\gamma$) exponent value of the power law estimated from the tail region of $P(k)$ in log-log plot differs between both cases with higher value for UDVG. Concentration vs degree (v-k) plots highlighted that hubs are related to the highest $PM10$ concentrations in VG while hubs is associated to the lowest concentrations in UDVG, i.e. probably the background atmosphere.
	
	For the overall, the multifractal analysis was then carried out using Rényi and singularity spectra for VG and UDVG. A clear difference of behavior between VG and UDVG is noticed with lower value of fractal object ($D_0$) for UDVG. Both spectra showed that multifractal degree is higher for VG than UDVG. Consequently, fluctuations for high concentrations are more significant than fluctuations for low concentrations. Both UDVG spectra shapes confirmed that this approach represents the background atmosphere due to the persistence of low $PM10$ concentrations related to marine aerosols and anthropogenic pollution in insular context.
	
	Thereafter, the same analysis was repeated according to African dust seasonality. This time, degree distribution analysis showed that the difference between VG and UDVG is reduced for the high season contrary to the low one. In VG frame, the multifractal degree is higher for the low season as expected. As regards UDVG frame, the multifractal degree is now higher for the high season. This opposite trend observed in UVDG is due to the increase of $PM10$ background atmosphere concentration from May to September. Indeed, contrary to the low season, there is a dusty background atmosphere in the high season due to the continuous alternation between African Easterly Waves and dust outbreaks.
	
	In conclusion, all these results pointed out that UDGV in complex network may be an efficient tool to perform the analysis of noise fluctuations in environmental time series. For the first time, UDVG frame is used outside of just identification of singularities, i.e. global behavior recognition. To precisely observe $PM10$ background atmosphere variation through the year, the monthly behavior of the degree distribution profile and the multifractal characteristics should be considered in future studies with more years.

\section*{Acknowledgements}

The authors are very grateful to the anonymous reviewers for their valuable comments and constructive suggestions, which helped us to improve substantially the quality of the paper. The authors would like to thank Guadeloupe air quality network (Gwad'Air) for providing air quality data.

\section*{Disclosure statement}

No potential conflict of interest was reported by the authors.

\section*{Funding}

The authors declare that they have not received any fund for the present paper. The paper is the sole work of the authors and is not a part/product of any project. 

\clearpage

\section*{References}
\bibliographystyle{model4-names}
\biboptions{authoryear}

\bibliography{VGUDVGPM10_Rev}

\begin{thebibliography}{96}
\expandafter\ifx\csname natexlab\endcsname\relax\def\natexlab#1{#1}\fi
\providecommand{\url}[1]{\texttt{#1}}
\providecommand{\href}[2]{#2}
\providecommand{\path}[1]{#1}
\providecommand{\DOIprefix}{doi:}
\providecommand{\ArXivprefix}{arXiv:}
\providecommand{\URLprefix}{URL: }
\providecommand{\Pubmedprefix}{pmid:}
\providecommand{\doi}[1]{\href{http://dx.doi.org/#1}{\path{#1}}}
\providecommand{\Pubmed}[1]{\href{pmid:#1}{\path{#1}}}
\providecommand{\bibinfo}[2]{#2}
\ifx\xfnm\undefined \def\xfnm[#1]{\unskip,\space#1}\fi
\bibitem[{Adams et~al.(2012)Adams, Prospero and Zhang}]{adams2012}
\bibinfo{author}{Adams\xfnm[ A.M.]}, \bibinfo{author}{Prospero\xfnm[ J.M.]},
  \bibinfo{author}{Zhang\xfnm[ C.]}.
\newblock \bibinfo{title}{{CALIPSO-derived three-dimensional structure of
  aerosol over the Atlantic Basin and adjacent continents}}.
\newblock \bibinfo{journal}{Journal of Climate}
  \bibinfo{year}{2012};\bibinfo{volume}{25}(\bibinfo{number}{19}):\bibinfo{pages}{6862--6879}.
\bibitem[{Bacry et~al.(1993)Bacry, Muzy and Arneodo}]{bacry1993}
\bibinfo{author}{Bacry\xfnm[ E.]}, \bibinfo{author}{Muzy\xfnm[ J.F.]},
  \bibinfo{author}{Arneodo\xfnm[ A.]}.
\newblock \bibinfo{title}{{Singularity spectrum of fractal signals from wavelet
  analysis: Exact results}}.
\newblock \bibinfo{journal}{Journal of statistical physics}
  \bibinfo{year}{1993};\bibinfo{volume}{70}(\bibinfo{number}{3-4}):\bibinfo{pages}{635--674}.
\bibitem[{Baranowski et~al.(2015)Baranowski, Krzyszczak, Slawinski, Hoffmann,
  Kozyra, Nier{\'o}bca, Siwek and Gluza}]{baranowski2015}
\bibinfo{author}{Baranowski\xfnm[ P.]}, \bibinfo{author}{Krzyszczak\xfnm[ J.]},
  \bibinfo{author}{Slawinski\xfnm[ C.]}, \bibinfo{author}{Hoffmann\xfnm[ H.]},
  \bibinfo{author}{Kozyra\xfnm[ J.]}, \bibinfo{author}{Nier{\'o}bca\xfnm[ A.]},
  \bibinfo{author}{Siwek\xfnm[ K.]}, \bibinfo{author}{Gluza\xfnm[ A.]}.
\newblock \bibinfo{title}{Multifractal analysis of meteorological time series
  to assess climate impacts}.
\newblock \bibinfo{journal}{Climate Research}
  \bibinfo{year}{2015};\bibinfo{volume}{65}:\bibinfo{pages}{39--52}.
\bibitem[{Block et~al.(1990)Block, Von~Bloh and Schellnhuber}]{block1990}
\bibinfo{author}{Block\xfnm[ A.]}, \bibinfo{author}{Von~Bloh\xfnm[ W.]},
  \bibinfo{author}{Schellnhuber\xfnm[ H.]}.
\newblock \bibinfo{title}{Efficient box-counting determination of generalized
  fractal dimensions}.
\newblock \bibinfo{journal}{Physical Review A}
  \bibinfo{year}{1990};\bibinfo{volume}{42}(\bibinfo{number}{4}):\bibinfo{pages}{1869}.
\bibitem[{Burpee(1972)}]{burpee1972}
\bibinfo{author}{Burpee\xfnm[ R.W.]}.
\newblock \bibinfo{title}{{The origin and structure of easterly waves in the
  lower troposphere of North Africa}}.
\newblock \bibinfo{journal}{Journal of the Atmospheric Sciences}
  \bibinfo{year}{1972};\bibinfo{volume}{29}(\bibinfo{number}{1}):\bibinfo{pages}{77--90}.
\bibitem[{Cadelis et~al.(2014)Cadelis, Tourres and Molinie}]{cadelis2014}
\bibinfo{author}{Cadelis\xfnm[ G.]}, \bibinfo{author}{Tourres\xfnm[ R.]},
  \bibinfo{author}{Molinie\xfnm[ J.]}.
\newblock \bibinfo{title}{Short-term effects of the particulate pollutants
  contained in saharan dust on the visits of children to the emergency
  department due to asthmatic conditions in {Guadeloupe} ({French Archipelago
  of the Caribbean})}.
\newblock \bibinfo{journal}{PloS one}
  \bibinfo{year}{2014};\bibinfo{volume}{9}:\bibinfo{pages}{e91136}.
\bibitem[{Cadelis et~al.(2013)Cadelis, Tourres, Molinie and
  Petit}]{cadelis2013}
\bibinfo{author}{Cadelis\xfnm[ G.]}, \bibinfo{author}{Tourres\xfnm[ R.]},
  \bibinfo{author}{Molinie\xfnm[ J.]}, \bibinfo{author}{Petit\xfnm[ R.]}.
\newblock \bibinfo{title}{Exacerbations d’asthme en guadeloupe et
  {\'e}ruption volcanique {\`a} montserrat (70 km de la guadeloupe)}.
\newblock \bibinfo{journal}{Revue des maladies respiratoires}
  \bibinfo{year}{2013};\bibinfo{volume}{30}(\bibinfo{number}{3}):\bibinfo{pages}{203--214}.
\bibitem[{Calif and Schmitt(2014)}]{calif2014}
\bibinfo{author}{Calif\xfnm[ R.]}, \bibinfo{author}{Schmitt\xfnm[ F.G.]}.
\newblock \bibinfo{title}{Multiscaling and joint multiscaling description of
  the atmospheric wind speed and the aggregate power output from a wind farm}.
\newblock \bibinfo{journal}{Nonlinear Processes in Geophysics}
  \bibinfo{year}{2014};\bibinfo{volume}{21}:\bibinfo{pages}{379--392}.
\bibitem[{Calif et~al.(2013)Calif, Schmitt, Huang and Soubdhan}]{calif2013}
\bibinfo{author}{Calif\xfnm[ R.]}, \bibinfo{author}{Schmitt\xfnm[ F.G.]},
  \bibinfo{author}{Huang\xfnm[ Y.]}, \bibinfo{author}{Soubdhan\xfnm[ T.]}.
\newblock \bibinfo{title}{Intermittency study of high frequency global solar
  radiation sequences under a tropical climate}.
\newblock \bibinfo{journal}{Solar Energy}
  \bibinfo{year}{2013};\bibinfo{volume}{98}:\bibinfo{pages}{349--365}.
\bibitem[{Caniego et~al.(2005)Caniego, Espejo, Mart{\i}n and
  San~Jos{\'e}}]{caniego2005}
\bibinfo{author}{Caniego\xfnm[ F.]}, \bibinfo{author}{Espejo\xfnm[ R.]},
  \bibinfo{author}{Mart{\i}n\xfnm[ M.]}, \bibinfo{author}{San~Jos{\'e}\xfnm[
  F.]}.
\newblock \bibinfo{title}{Multifractal scaling of soil spatial variability}.
\newblock \bibinfo{journal}{Ecological Modelling}
  \bibinfo{year}{2005};\bibinfo{volume}{182}(\bibinfo{number}{3-4}):\bibinfo{pages}{291--303}.
\bibitem[{Carlson(1969)}]{carlson1969}
\bibinfo{author}{Carlson\xfnm[ T.]}.
\newblock \bibinfo{title}{{Synoptic histories on African disturbances and their
  progress over the tropical Atlantic}}.
\newblock \bibinfo{journal}{Mon Weather Rev}
  \bibinfo{year}{1969};\bibinfo{volume}{97}:\bibinfo{pages}{256--276}.
\bibitem[{Carmona-Cabezas et~al.(2019{\natexlab{a}})Carmona-Cabezas,
  Ariza-Villaverde, Guti{\'e}rrez~de Rav{\'e} and
  Jim{\'e}nez-Hornero}]{carmona2019b}
\bibinfo{author}{Carmona-Cabezas\xfnm[ R.]},
  \bibinfo{author}{Ariza-Villaverde\xfnm[ A.B.]},
  \bibinfo{author}{Guti{\'e}rrez~de Rav{\'e}\xfnm[ E.]},
  \bibinfo{author}{Jim{\'e}nez-Hornero\xfnm[ F.J.]}.
\newblock \bibinfo{title}{{Visibility graphs of ground-level ozone time series:
  A multifractal analysis}}.
\newblock \bibinfo{journal}{Science of The Total Environment}
  \bibinfo{year}{2019}{\natexlab{a}};\bibinfo{volume}{661}:\bibinfo{pages}{138--147}.
\bibitem[{Carmona-Cabezas et~al.(2019{\natexlab{b}})Carmona-Cabezas,
  G{\'o}mez-G{\'o}mez, Ariza-Villaverde, Guti{\'e}rrez~de Rav{\'e} and
  Jim{\'e}nez-Hornero}]{carmona2019a}
\bibinfo{author}{Carmona-Cabezas\xfnm[ R.]},
  \bibinfo{author}{G{\'o}mez-G{\'o}mez\xfnm[ J.]},
  \bibinfo{author}{Ariza-Villaverde\xfnm[ A.B.]},
  \bibinfo{author}{Guti{\'e}rrez~de Rav{\'e}\xfnm[ E.]},
  \bibinfo{author}{Jim{\'e}nez-Hornero\xfnm[ F.J.]}.
\newblock \bibinfo{title}{{Can complex networks describe the urban and rural
  tropospheric $O_3$ dynamics?}}
\newblock \bibinfo{journal}{Chemosphere}
  \bibinfo{year}{2019}{\natexlab{b}};\bibinfo{volume}{230}:\bibinfo{pages}{59--66}.
\bibitem[{Carmona-Cabezas et~al.(2020)Carmona-Cabezas, G{\'o}mez-G{\'o}mez,
  Guti{\'e}rrez~de Rav{\'e}, S{\'a}nchez-L{\'o}pez, Serrano and
  Jim{\'e}nez-Hornero}]{carmona2020a}
\bibinfo{author}{Carmona-Cabezas\xfnm[ R.]},
  \bibinfo{author}{G{\'o}mez-G{\'o}mez\xfnm[ J.]},
  \bibinfo{author}{Guti{\'e}rrez~de Rav{\'e}\xfnm[ E.]},
  \bibinfo{author}{S{\'a}nchez-L{\'o}pez\xfnm[ E.]},
  \bibinfo{author}{Serrano\xfnm[ J.]},
  \bibinfo{author}{Jim{\'e}nez-Hornero\xfnm[ F.J.]}.
\newblock \bibinfo{title}{Improving graph-based detection of singular events
  for photochemical smog agents}.
\newblock \bibinfo{journal}{Chemosphere}
  \bibinfo{year}{2020};:\bibinfo{pages}{126660}.
\bibitem[{Chhabra et~al.(1989)Chhabra, Meneveau, Jensen and
  Sreenivasan}]{chhabra1989}
\bibinfo{author}{Chhabra\xfnm[ A.B.]}, \bibinfo{author}{Meneveau\xfnm[ C.]},
  \bibinfo{author}{Jensen\xfnm[ R.V.]}, \bibinfo{author}{Sreenivasan\xfnm[
  K.]}.
\newblock \bibinfo{title}{Direct determination of the f ($\alpha$) singularity
  spectrum and its application to fully developed turbulence}.
\newblock \bibinfo{journal}{Physical Review A}
  \bibinfo{year}{1989};\bibinfo{volume}{40}(\bibinfo{number}{9}):\bibinfo{pages}{5284}.
\bibitem[{Clergue et~al.(2015)Clergue, Dellinger, Buss, Gaillardet, Benedetti
  and Dessert}]{clergue2015}
\bibinfo{author}{Clergue\xfnm[ C.]}, \bibinfo{author}{Dellinger\xfnm[ M.]},
  \bibinfo{author}{Buss\xfnm[ H.]}, \bibinfo{author}{Gaillardet\xfnm[ J.]},
  \bibinfo{author}{Benedetti\xfnm[ M.]}, \bibinfo{author}{Dessert\xfnm[ C.]}.
\newblock \bibinfo{title}{{Influence of atmospheric deposits and secondary
  minerals on Li isotopes budget in a highly weathered catchment, Guadeloupe
  (Lesser Antilles)}}.
\newblock \bibinfo{journal}{Chemical Geology}
  \bibinfo{year}{2015};\bibinfo{volume}{414}:\bibinfo{pages}{28--41}.
\bibitem[{D'Alessandro et~al.(2003)D'Alessandro, Lucarelli, Mand{\`o},
  Marcazzan, Nava, Prati, Valli, Vecchi and Zucchiatti}]{alessandro2003}
\bibinfo{author}{D'Alessandro\xfnm[ A.]}, \bibinfo{author}{Lucarelli\xfnm[
  F.]}, \bibinfo{author}{Mand{\`o}\xfnm[ P.]}, \bibinfo{author}{Marcazzan\xfnm[
  G.]}, \bibinfo{author}{Nava\xfnm[ S.]}, \bibinfo{author}{Prati\xfnm[ P.]},
  \bibinfo{author}{Valli\xfnm[ G.]}, \bibinfo{author}{Vecchi\xfnm[ R.]},
  \bibinfo{author}{Zucchiatti\xfnm[ A.]}.
\newblock \bibinfo{title}{{Hourly elemental composition and sources
  identification of fine and coarse PM10 particulate matter in four Italian
  towns}}.
\newblock \bibinfo{journal}{Journal of Aerosol Science}
  \bibinfo{year}{2003};\bibinfo{volume}{34}(\bibinfo{number}{2}):\bibinfo{pages}{243--259}.
\bibitem[{Dong et~al.(2017)Dong, Wang and Li}]{dong2017}
\bibinfo{author}{Dong\xfnm[ Q.]}, \bibinfo{author}{Wang\xfnm[ Y.]},
  \bibinfo{author}{Li\xfnm[ P.]}.
\newblock \bibinfo{title}{Multifractal behavior of an air pollutant time series
  and the relevance to the predictability}.
\newblock \bibinfo{journal}{Environmental Pollution}
  \bibinfo{year}{2017};\bibinfo{volume}{222}:\bibinfo{pages}{444--457}.
\bibitem[{Euphrasie-Clotilde et~al.(2020)Euphrasie-Clotilde, Plocoste,
  Feuillard, Velasco-Merino, Mateos, Toledano, Brute, Bassette and
  Gobinddass}]{euphrasie2020}
\bibinfo{author}{Euphrasie-Clotilde\xfnm[ L.]}, \bibinfo{author}{Plocoste\xfnm[
  T.]}, \bibinfo{author}{Feuillard\xfnm[ T.]},
  \bibinfo{author}{Velasco-Merino\xfnm[ C.]}, \bibinfo{author}{Mateos\xfnm[
  D.]}, \bibinfo{author}{Toledano\xfnm[ C.]}, \bibinfo{author}{Brute\xfnm[
  F.N.]}, \bibinfo{author}{Bassette\xfnm[ C.]},
  \bibinfo{author}{Gobinddass\xfnm[ M.]}.
\newblock \bibinfo{title}{{Assessment of a new detection threshold for PM10
  concentrations linked to African dust events in the Caribbean Basin}}.
\newblock \bibinfo{journal}{Atmospheric Environment}
  \bibinfo{year}{2020};\bibinfo{volume}{224}:\bibinfo{pages}{117354}.
\bibitem[{Feder(1988)}]{feder1988}
\bibinfo{author}{Feder\xfnm[ J.]}.
\newblock \bibinfo{title}{{Fractals (Physics of Solids and Liquids)}}.
\newblock \bibinfo{edition}{1988th} ed.
\newblock \bibinfo{publisher}{Springer, Boston}, \bibinfo{year}{1988}.
\bibitem[{Feng et~al.(2019)Feng, Li, Wang, Van Halm-Lutterodt, An, Liu, Liu,
  Wang and Guo}]{feng2019}
\bibinfo{author}{Feng\xfnm[ W.]}, \bibinfo{author}{Li\xfnm[ H.]},
  \bibinfo{author}{Wang\xfnm[ S.]}, \bibinfo{author}{Van Halm-Lutterodt\xfnm[
  N.]}, \bibinfo{author}{An\xfnm[ J.]}, \bibinfo{author}{Liu\xfnm[ Y.]},
  \bibinfo{author}{Liu\xfnm[ M.]}, \bibinfo{author}{Wang\xfnm[ X.]},
  \bibinfo{author}{Guo\xfnm[ X.]}.
\newblock \bibinfo{title}{{Short-term PM10 and emergency department admissions
  for selective cardiovascular and respiratory diseases in Beijing, China}}.
\newblock \bibinfo{journal}{Science of The Total Environment}
  \bibinfo{year}{2019};\bibinfo{volume}{657}:\bibinfo{pages}{213--221}.
\bibitem[{Gao et~al.(2016)Gao, Wang, Huang, Zhou, Lu, Shi and Liu}]{gao2016}
\bibinfo{author}{Gao\xfnm[ S.]}, \bibinfo{author}{Wang\xfnm[ Y.]},
  \bibinfo{author}{Huang\xfnm[ Y.]}, \bibinfo{author}{Zhou\xfnm[ Q.]},
  \bibinfo{author}{Lu\xfnm[ Z.]}, \bibinfo{author}{Shi\xfnm[ X.]},
  \bibinfo{author}{Liu\xfnm[ Y.]}.
\newblock \bibinfo{title}{Spatial statistics of atmospheric particulate matter
  in {China}}.
\newblock \bibinfo{journal}{Atmospheric Environment}
  \bibinfo{year}{2016};\bibinfo{volume}{134}:\bibinfo{pages}{162--167}.
\bibitem[{Grivas and Chaloulakou(2006)}]{grivas2006}
\bibinfo{author}{Grivas\xfnm[ G.]}, \bibinfo{author}{Chaloulakou\xfnm[ A.]}.
\newblock \bibinfo{title}{{Artificial neural network models for prediction of
  PM10 hourly concentrations, in the Greater Area of Athens, Greece}}.
\newblock \bibinfo{journal}{Atmospheric environment}
  \bibinfo{year}{2006};\bibinfo{volume}{40}(\bibinfo{number}{7}):\bibinfo{pages}{1216--1229}.
\bibitem[{Gurung et~al.(2017)Gurung, Son and Bell}]{gurung2017}
\bibinfo{author}{Gurung\xfnm[ A.]}, \bibinfo{author}{Son\xfnm[ J.Y.]},
  \bibinfo{author}{Bell\xfnm[ M.L.]}.
\newblock \bibinfo{title}{{Particulate matter and risk of hospital admission in
  the Kathmandu Valley, Nepal: a case-crossover study}}.
\newblock \bibinfo{journal}{American journal of epidemiology}
  \bibinfo{year}{2017};\bibinfo{volume}{186}(\bibinfo{number}{5}):\bibinfo{pages}{573--580}.
\bibitem[{Halsey et~al.(1986)Halsey, Jensen, Kadanoff, Procaccia and
  Shraiman}]{halsey1986}
\bibinfo{author}{Halsey\xfnm[ T.C.]}, \bibinfo{author}{Jensen\xfnm[ M.H.]},
  \bibinfo{author}{Kadanoff\xfnm[ L.P.]}, \bibinfo{author}{Procaccia\xfnm[
  I.]}, \bibinfo{author}{Shraiman\xfnm[ B.I.]}.
\newblock \bibinfo{title}{{Fractal measures and their singularities: The
  characterization of strange sets}}.
\newblock \bibinfo{journal}{Physical Review A}
  \bibinfo{year}{1986};\bibinfo{volume}{33}(\bibinfo{number}{2}):\bibinfo{pages}{1141}.
\bibitem[{Harte(2001)}]{harte2001}
\bibinfo{author}{Harte\xfnm[ D.]}.
\newblock \bibinfo{title}{Multifractals: theory and applications}.
\newblock \bibinfo{publisher}{CRC Press}, \bibinfo{year}{2001}.
\bibitem[{Ho et~al.(2004)Ho, Juang, Liao, Wang, Lee, Hsu, Yang and Yu}]{ho2004}
\bibinfo{author}{Ho\xfnm[ D.S.]}, \bibinfo{author}{Juang\xfnm[ L.C.]},
  \bibinfo{author}{Liao\xfnm[ Y.Y.]}, \bibinfo{author}{Wang\xfnm[ C.C.]},
  \bibinfo{author}{Lee\xfnm[ C.K.]}, \bibinfo{author}{Hsu\xfnm[ T.C.]},
  \bibinfo{author}{Yang\xfnm[ S.Y.]}, \bibinfo{author}{Yu\xfnm[ C.C.]}.
\newblock \bibinfo{title}{{The temporal variations of PM10 concentration in
  Taipei: a fractal approach}}.
\newblock \bibinfo{journal}{Aerosol and Air Quality Research}
  \bibinfo{year}{2004};\bibinfo{volume}{4}(\bibinfo{number}{1}):\bibinfo{pages}{38--55}.
\bibitem[{Huang et~al.(2010)Huang, Zhang and Prospero}]{huang2010}
\bibinfo{author}{Huang\xfnm[ J.]}, \bibinfo{author}{Zhang\xfnm[ C.]},
  \bibinfo{author}{Prospero\xfnm[ J.M.]}.
\newblock \bibinfo{title}{{African dust outbreaks: A satellite perspective of
  temporal and spatial variability over the tropical Atlantic Ocean}}.
\newblock \bibinfo{journal}{Journal of Geophysical Research: Atmospheres}
  \bibinfo{year}{2010};\bibinfo{volume}{115}(\bibinfo{number}{D5}).
\bibitem[{Jickells et~al.(2005)Jickells, An, Andersen, Baker, Bergametti,
  Brooks, Cao, Boyd, Duce, Hunter, Kawahata, Kubilay, LaRoche, Liss, Mahowald,
  Prospero, Ridgwell, Tegen and Prato~Torres}]{jickells2005}
\bibinfo{author}{Jickells\xfnm[ T.]}, \bibinfo{author}{An\xfnm[ Z.]},
  \bibinfo{author}{Andersen\xfnm[ K.K.]}, \bibinfo{author}{Baker\xfnm[ A.]},
  \bibinfo{author}{Bergametti\xfnm[ G.]}, \bibinfo{author}{Brooks\xfnm[ N.]},
  \bibinfo{author}{Cao\xfnm[ J.]}, \bibinfo{author}{Boyd\xfnm[ P.]},
  \bibinfo{author}{Duce\xfnm[ R.]}, \bibinfo{author}{Hunter\xfnm[ K.]},
  \bibinfo{author}{Kawahata\xfnm[ H.]}, \bibinfo{author}{Kubilay\xfnm[ N.]},
  \bibinfo{author}{LaRoche\xfnm[ J.]}, \bibinfo{author}{Liss\xfnm[ P.]},
  \bibinfo{author}{Mahowald\xfnm[ N.]}, \bibinfo{author}{Prospero\xfnm[ J.]},
  \bibinfo{author}{Ridgwell\xfnm[ A.]}, \bibinfo{author}{Tegen\xfnm[ I.]},
  \bibinfo{author}{Prato~Torres\xfnm[ R.]}.
\newblock \bibinfo{title}{Global iron connections between desert dust, ocean
  biogeochemistry, and climate}.
\newblock \bibinfo{journal}{science}
  \bibinfo{year}{2005};\bibinfo{volume}{308}(\bibinfo{number}{5718}):\bibinfo{pages}{67--71}.
\bibitem[{Jim{\'e}nez-Hornero et~al.(2011)Jim{\'e}nez-Hornero,
  Pav{\'o}n-Dom{\'\i}nguez, de~Rav{\'e} and Ariza-Villaverde}]{jimenez2011}
\bibinfo{author}{Jim{\'e}nez-Hornero\xfnm[ F.]},
  \bibinfo{author}{Pav{\'o}n-Dom{\'\i}nguez\xfnm[ P.]},
  \bibinfo{author}{de~Rav{\'e}\xfnm[ E.G.]},
  \bibinfo{author}{Ariza-Villaverde\xfnm[ A.]}.
\newblock \bibinfo{title}{Joint multifractal description of the relationship
  between wind patterns and land surface air temperature}.
\newblock \bibinfo{journal}{Atmospheric research}
  \bibinfo{year}{2011};\bibinfo{volume}{99}(\bibinfo{number}{3-4}):\bibinfo{pages}{366--376}.
\bibitem[{Jury(2017)}]{jury2017}
\bibinfo{author}{Jury\xfnm[ M.R.]}.
\newblock \bibinfo{title}{{Caribbean Air Chemistry and Dispersion Conditions}}.
\newblock \bibinfo{journal}{Atmosphere}
  \bibinfo{year}{2017};\bibinfo{volume}{8}(\bibinfo{number}{8}):\bibinfo{pages}{151}.
\bibitem[{Karyampudi and Carlson(1988)}]{karyampudi1988}
\bibinfo{author}{Karyampudi\xfnm[ V.M.]}, \bibinfo{author}{Carlson\xfnm[
  T.N.]}.
\newblock \bibinfo{title}{{Analysis and numerical simulations of the Saharan
  air layer and its effect on easterly wave disturbances}}.
\newblock \bibinfo{journal}{Journal of the Atmospheric Sciences}
  \bibinfo{year}{1988};\bibinfo{volume}{45}(\bibinfo{number}{21}):\bibinfo{pages}{3102--3136}.
\bibitem[{Karyampudi et~al.(1999)Karyampudi, Palm, Reagen, Fang, Grant, Hoff,
  Moulin, Pierce, Torres, Browell and Melfi}]{karyampudi1999}
\bibinfo{author}{Karyampudi\xfnm[ V.M.]}, \bibinfo{author}{Palm\xfnm[ S.P.]},
  \bibinfo{author}{Reagen\xfnm[ J.A.]}, \bibinfo{author}{Fang\xfnm[ H.]},
  \bibinfo{author}{Grant\xfnm[ W.B.]}, \bibinfo{author}{Hoff\xfnm[ R.M.]},
  \bibinfo{author}{Moulin\xfnm[ C.]}, \bibinfo{author}{Pierce\xfnm[ H.F.]},
  \bibinfo{author}{Torres\xfnm[ O.]}, \bibinfo{author}{Browell\xfnm[ E.V.]},
  \bibinfo{author}{Melfi\xfnm[ S.H.]}.
\newblock \bibinfo{title}{{Validation of the Saharan dust plume conceptual
  model using lidar, Meteosat, and ECMWF data}}.
\newblock \bibinfo{journal}{Bulletin of the American Meteorological Society}
  \bibinfo{year}{1999};\bibinfo{volume}{80}(\bibinfo{number}{6}):\bibinfo{pages}{1045--1076}.
\bibitem[{Kelty-Stephen et~al.(2013)Kelty-Stephen, Palatinus, Saltzman and
  Dixon}]{kelty2013}
\bibinfo{author}{Kelty-Stephen\xfnm[ D.G.]}, \bibinfo{author}{Palatinus\xfnm[
  K.]}, \bibinfo{author}{Saltzman\xfnm[ E.]}, \bibinfo{author}{Dixon\xfnm[
  J.A.]}.
\newblock \bibinfo{title}{A tutorial on multifractality, cascades, and
  interactivity for empirical time series in ecological science}.
\newblock \bibinfo{journal}{Ecological Psychology}
  \bibinfo{year}{2013};\bibinfo{volume}{25}(\bibinfo{number}{1}):\bibinfo{pages}{1--62}.
\bibitem[{Kravchenko et~al.(1999)Kravchenko, Boast and
  Bullock}]{kravchenko1999}
\bibinfo{author}{Kravchenko\xfnm[ A.N.]}, \bibinfo{author}{Boast\xfnm[ C.W.]},
  \bibinfo{author}{Bullock\xfnm[ D.G.]}.
\newblock \bibinfo{title}{Multifractal analysis of soil spatial variability}.
\newblock \bibinfo{journal}{Agronomy Journal}
  \bibinfo{year}{1999};\bibinfo{volume}{91}(\bibinfo{number}{6}):\bibinfo{pages}{1033--1041}.
\bibitem[{Lacasa et~al.(2008)Lacasa, Luque, Ballesteros, Luque and
  Nuno}]{lacasa2008}
\bibinfo{author}{Lacasa\xfnm[ L.]}, \bibinfo{author}{Luque\xfnm[ B.]},
  \bibinfo{author}{Ballesteros\xfnm[ F.]}, \bibinfo{author}{Luque\xfnm[ J.]},
  \bibinfo{author}{Nuno\xfnm[ J.C.]}.
\newblock \bibinfo{title}{{From time series to complex networks: The visibility
  graph}}.
\newblock \bibinfo{journal}{Proceedings of the National Academy of Sciences}
  \bibinfo{year}{2008};\bibinfo{volume}{105}(\bibinfo{number}{13}):\bibinfo{pages}{4972--4975}.
\bibitem[{Lacasa et~al.(2009)Lacasa, Luque, Luque and Nuno}]{lacasa2009}
\bibinfo{author}{Lacasa\xfnm[ L.]}, \bibinfo{author}{Luque\xfnm[ B.]},
  \bibinfo{author}{Luque\xfnm[ J.]}, \bibinfo{author}{Nuno\xfnm[ J.C.]}.
\newblock \bibinfo{title}{{The visibility graph: A new method for estimating
  the Hurst exponent of fractional Brownian motion}}.
\newblock \bibinfo{journal}{EPL (Europhysics Letters)}
  \bibinfo{year}{2009};\bibinfo{volume}{86}(\bibinfo{number}{3}):\bibinfo{pages}{30001}.
\bibitem[{Lacasa et~al.(2015)Lacasa, Nicosia and Latora}]{lacasa2015}
\bibinfo{author}{Lacasa\xfnm[ L.]}, \bibinfo{author}{Nicosia\xfnm[ V.]},
  \bibinfo{author}{Latora\xfnm[ V.]}.
\newblock \bibinfo{title}{Network structure of multivariate time series}.
\newblock \bibinfo{journal}{Scientific reports}
  \bibinfo{year}{2015};\bibinfo{volume}{5}:\bibinfo{pages}{15508}.
\bibitem[{Lacasa and Toral(2010)}]{lacasa2010}
\bibinfo{author}{Lacasa\xfnm[ L.]}, \bibinfo{author}{Toral\xfnm[ R.]}.
\newblock \bibinfo{title}{Description of stochastic and chaotic series using
  visibility graphs}.
\newblock \bibinfo{journal}{Physical Review E}
  \bibinfo{year}{2010};\bibinfo{volume}{82}(\bibinfo{number}{3}):\bibinfo{pages}{036120}.
\bibitem[{Lee(2002)}]{lee2002}
\bibinfo{author}{Lee\xfnm[ C.K.]}.
\newblock \bibinfo{title}{Multifractal characteristics in air pollutant
  concentration time series}.
\newblock \bibinfo{journal}{Water, Air, and Soil Pollution}
  \bibinfo{year}{2002};\bibinfo{volume}{135}(\bibinfo{number}{1-4}):\bibinfo{pages}{389--409}.
\bibitem[{Ling and van Eeden(2009)}]{ling2009}
\bibinfo{author}{Ling\xfnm[ S.H.]}, \bibinfo{author}{van Eeden\xfnm[ S.F.]}.
\newblock \bibinfo{title}{Particulate matter air pollution exposure: role in
  the development and exacerbation of chronic obstructive pulmonary disease}.
\newblock \bibinfo{journal}{International journal of chronic obstructive
  pulmonary disease}
  \bibinfo{year}{2009};\bibinfo{volume}{4}:\bibinfo{pages}{233}.
\bibitem[{Liu et~al.(2015{\natexlab{a}})Liu, Yu and Anh}]{liu2015b}
\bibinfo{author}{Liu\xfnm[ J.L.]}, \bibinfo{author}{Yu\xfnm[ Z.G.]},
  \bibinfo{author}{Anh\xfnm[ V.]}.
\newblock \bibinfo{title}{Determination of multifractal dimensions of complex
  networks by means of the sandbox algorithm}.
\newblock \bibinfo{journal}{Chaos: An Interdisciplinary Journal of Nonlinear
  Science}
  \bibinfo{year}{2015}{\natexlab{a}};\bibinfo{volume}{25}(\bibinfo{number}{2}):\bibinfo{pages}{023103}.
\bibitem[{Liu et~al.(2015{\natexlab{b}})Liu, Wang and Zhu}]{liu2015a}
\bibinfo{author}{Liu\xfnm[ Z.]}, \bibinfo{author}{Wang\xfnm[ L.]},
  \bibinfo{author}{Zhu\xfnm[ H.]}.
\newblock \bibinfo{title}{A time--scaling property of air pollution indices: a
  case study of {Shanghai}, {China}}.
\newblock \bibinfo{journal}{Atmospheric Pollution Research}
  \bibinfo{year}{2015}{\natexlab{b}};\bibinfo{volume}{6}:\bibinfo{pages}{886--892}.
\bibitem[{Lyra and Tsallis(1998)}]{lyra1998}
\bibinfo{author}{Lyra\xfnm[ M.]}, \bibinfo{author}{Tsallis\xfnm[ C.]}.
\newblock \bibinfo{title}{Nonextensivity and multifractality in low-dimensional
  dissipative systems}.
\newblock \bibinfo{journal}{Physical review letters}
  \bibinfo{year}{1998};\bibinfo{volume}{80}(\bibinfo{number}{1}):\bibinfo{pages}{53}.
\bibitem[{Mach et~al.(1995)Mach, Mas and Sagu{\'e}s}]{mach1995}
\bibinfo{author}{Mach\xfnm[ J.]}, \bibinfo{author}{Mas\xfnm[ F.]},
  \bibinfo{author}{Sagu{\'e}s\xfnm[ F.]}.
\newblock \bibinfo{title}{Two representations in multifractal analysis}.
\newblock \bibinfo{journal}{Journal of Physics A: Mathematical and General}
  \bibinfo{year}{1995};\bibinfo{volume}{28}(\bibinfo{number}{19}):\bibinfo{pages}{5607}.
\bibitem[{Mahowald et~al.(2014)Mahowald, Albani, Kok, Engelstaeder, Scanza,
  Ward and Flanner}]{mahowald2014}
\bibinfo{author}{Mahowald\xfnm[ N.]}, \bibinfo{author}{Albani\xfnm[ S.]},
  \bibinfo{author}{Kok\xfnm[ J.F.]}, \bibinfo{author}{Engelstaeder\xfnm[ S.]},
  \bibinfo{author}{Scanza\xfnm[ R.]}, \bibinfo{author}{Ward\xfnm[ D.S.]},
  \bibinfo{author}{Flanner\xfnm[ M.G.]}.
\newblock \bibinfo{title}{The size distribution of desert dust aerosols and its
  impact on the {Earth} system}.
\newblock \bibinfo{journal}{Aeolian Research}
  \bibinfo{year}{2014};\bibinfo{volume}{15}:\bibinfo{pages}{53--71}.
\bibitem[{Maleki et~al.(2016)Maleki, Sorooshian, Goudarzi, Nikfal and
  Baneshi}]{maleki2016}
\bibinfo{author}{Maleki\xfnm[ H.]}, \bibinfo{author}{Sorooshian\xfnm[ A.]},
  \bibinfo{author}{Goudarzi\xfnm[ G.]}, \bibinfo{author}{Nikfal\xfnm[ A.]},
  \bibinfo{author}{Baneshi\xfnm[ M.M.]}.
\newblock \bibinfo{title}{{Temporal profile of PM10 and associated health
  effects in one of the most polluted cities of the world (Ahvaz, Iran) between
  2009 and 2014}}.
\newblock \bibinfo{journal}{Aeolian research}
  \bibinfo{year}{2016};\bibinfo{volume}{22}:\bibinfo{pages}{135--140}.
\bibitem[{Mali et~al.(2018)Mali, Manna, Mukhopadhyay, Haldar and
  Singh}]{mali2018}
\bibinfo{author}{Mali\xfnm[ P.]}, \bibinfo{author}{Manna\xfnm[ S.]},
  \bibinfo{author}{Mukhopadhyay\xfnm[ A.]}, \bibinfo{author}{Haldar\xfnm[ P.]},
  \bibinfo{author}{Singh\xfnm[ G.]}.
\newblock \bibinfo{title}{Multifractal analysis of multiparticle emission data
  in the framework of visibility graph and sandbox algorithm}.
\newblock \bibinfo{journal}{Physica A: Statistical Mechanics and its
  Applications}
  \bibinfo{year}{2018};\bibinfo{volume}{493}:\bibinfo{pages}{253--266}.
\bibitem[{Mandelbrot(1982)}]{mandelbrot1982}
\bibinfo{author}{Mandelbrot\xfnm[ B.]}.
\newblock \bibinfo{title}{The Fractal Geometry of Nature}.
\newblock \bibinfo{publisher}{Freeman and Company, New York},
  \bibinfo{year}{1982}.
\bibitem[{Mandelbrot(1974)}]{mandelbrot1974}
\bibinfo{author}{Mandelbrot\xfnm[ B.B.]}.
\newblock \bibinfo{title}{Intermittent turbulence in self-similar cascades:
  divergence of high moments and dimension of the carrier}.
\newblock \bibinfo{journal}{Journal of fluid Mechanics}
  \bibinfo{year}{1974};\bibinfo{volume}{62}(\bibinfo{number}{2}):\bibinfo{pages}{331--358}.
\bibitem[{Martin et~al.(1991)Martin, Gordon and Fitzwater}]{martin1991}
\bibinfo{author}{Martin\xfnm[ J.H.]}, \bibinfo{author}{Gordon\xfnm[ M.]},
  \bibinfo{author}{Fitzwater\xfnm[ S.E.]}.
\newblock \bibinfo{title}{The case for iron}.
\newblock \bibinfo{journal}{Limnology and Oceanography}
  \bibinfo{year}{1991};\bibinfo{volume}{36}(\bibinfo{number}{8}):\bibinfo{pages}{1793--1802}.
\bibitem[{Momtazan et~al.(2019)Momtazan, Geravandi, Rastegarimehr, Valipour,
  Ranjbarzadeh, Yari, Dobaradaran, Bostan, Farhadi, Darabi, Khaniabadi and
  Mohammadi}]{momtazan2019}
\bibinfo{author}{Momtazan\xfnm[ M.]}, \bibinfo{author}{Geravandi\xfnm[ S.]},
  \bibinfo{author}{Rastegarimehr\xfnm[ B.]}, \bibinfo{author}{Valipour\xfnm[
  A.]}, \bibinfo{author}{Ranjbarzadeh\xfnm[ A.]}, \bibinfo{author}{Yari\xfnm[
  A.R.]}, \bibinfo{author}{Dobaradaran\xfnm[ S.]},
  \bibinfo{author}{Bostan\xfnm[ H.]}, \bibinfo{author}{Farhadi\xfnm[ M.]},
  \bibinfo{author}{Darabi\xfnm[ F.]}, \bibinfo{author}{Khaniabadi\xfnm[ Y.O.]},
  \bibinfo{author}{Mohammadi\xfnm[ M.J.]}.
\newblock \bibinfo{title}{{An investigation of particulate matter and relevant
  cardiovascular risks in Abadan and Khorramshahr in 2014--2016}}.
\newblock \bibinfo{journal}{Toxin reviews}
  \bibinfo{year}{2019};\bibinfo{volume}{38}(\bibinfo{number}{4}):\bibinfo{pages}{290--297}.
\bibitem[{Moreno et~al.(2006)Moreno, Querol, Castillo, Alastuey, Cuevas,
  Herrmann, Mounkaila, Elvira and Gibbons}]{moreno2006}
\bibinfo{author}{Moreno\xfnm[ T.]}, \bibinfo{author}{Querol\xfnm[ X.]},
  \bibinfo{author}{Castillo\xfnm[ S.]}, \bibinfo{author}{Alastuey\xfnm[ A.]},
  \bibinfo{author}{Cuevas\xfnm[ E.]}, \bibinfo{author}{Herrmann\xfnm[ L.]},
  \bibinfo{author}{Mounkaila\xfnm[ M.]}, \bibinfo{author}{Elvira\xfnm[ J.]},
  \bibinfo{author}{Gibbons\xfnm[ W.]}.
\newblock \bibinfo{title}{{Geochemical variations in aeolian mineral particles
  from the Sahara--Sahel Dust Corridor}}.
\newblock \bibinfo{journal}{Chemosphere}
  \bibinfo{year}{2006};\bibinfo{volume}{65}(\bibinfo{number}{2}):\bibinfo{pages}{261--270}.
\bibitem[{Muzy et~al.(1993)Muzy, Bacry and Arneodo}]{muzy1993}
\bibinfo{author}{Muzy\xfnm[ J.F.]}, \bibinfo{author}{Bacry\xfnm[ E.]},
  \bibinfo{author}{Arneodo\xfnm[ A.]}.
\newblock \bibinfo{title}{{Multifractal formalism for fractal signals: The
  structure-function approach versus the wavelet-transform modulus-maxima
  method}}.
\newblock \bibinfo{journal}{Physical review E}
  \bibinfo{year}{1993};\bibinfo{volume}{47}(\bibinfo{number}{2}):\bibinfo{pages}{875}.
\bibitem[{Okin et~al.(2008)Okin, Mladenov, Wang, Cassel, Caylor, Ringrose and
  Macko}]{okin2008}
\bibinfo{author}{Okin\xfnm[ G.]}, \bibinfo{author}{Mladenov\xfnm[ N.]},
  \bibinfo{author}{Wang\xfnm[ L.]}, \bibinfo{author}{Cassel\xfnm[ D.]},
  \bibinfo{author}{Caylor\xfnm[ K.]}, \bibinfo{author}{Ringrose\xfnm[ S.]},
  \bibinfo{author}{Macko\xfnm[ S.]}.
\newblock \bibinfo{title}{{Spatial patterns of soil nutrients in two southern
  African savannas}}.
\newblock \bibinfo{journal}{Journal of Geophysical Research: Biogeosciences}
  \bibinfo{year}{2008};\bibinfo{volume}{113}(\bibinfo{number}{G2}).
\bibitem[{Olsen(1995)}]{olsen1995}
\bibinfo{author}{Olsen\xfnm[ L.]}.
\newblock \bibinfo{title}{A multifractal formalism}.
\newblock \bibinfo{journal}{Advances in mathematics}
  \bibinfo{year}{1995};\bibinfo{volume}{116}(\bibinfo{number}{1}):\bibinfo{pages}{82--196}.
\bibitem[{Painter et~al.(2007)Painter, Barrett, Landry, Neff, Cassidy,
  Lawrence, McBride and Farmer}]{painter2007}
\bibinfo{author}{Painter\xfnm[ T.H.]}, \bibinfo{author}{Barrett\xfnm[ A.P.]},
  \bibinfo{author}{Landry\xfnm[ C.C.]}, \bibinfo{author}{Neff\xfnm[ J.C.]},
  \bibinfo{author}{Cassidy\xfnm[ M.P.]}, \bibinfo{author}{Lawrence\xfnm[
  C.R.]}, \bibinfo{author}{McBride\xfnm[ K.E.]}, \bibinfo{author}{Farmer\xfnm[
  G.L.]}.
\newblock \bibinfo{title}{Impact of disturbed desert soils on duration of
  mountain snow cover}.
\newblock \bibinfo{journal}{Geophysical Research Letters}
  \bibinfo{year}{2007};\bibinfo{volume}{34}(\bibinfo{number}{12}).
\bibitem[{Paschalidou et~al.(2011)Paschalidou, Karakitsios, Kleanthous and
  Kassomenos}]{paschalidou2011}
\bibinfo{author}{Paschalidou\xfnm[ A.K.]}, \bibinfo{author}{Karakitsios\xfnm[
  S.]}, \bibinfo{author}{Kleanthous\xfnm[ S.]},
  \bibinfo{author}{Kassomenos\xfnm[ P.A.]}.
\newblock \bibinfo{title}{{Forecasting hourly PM10 concentration in Cyprus
  through artificial neural networks and multiple regression models:
  implications to local environmental management}}.
\newblock \bibinfo{journal}{Environmental Science and Pollution Research}
  \bibinfo{year}{2011};\bibinfo{volume}{18}(\bibinfo{number}{2}):\bibinfo{pages}{316--327}.
\bibitem[{Petit et~al.(2005)Petit, Legrand, Jankowiak, Molini{\'e}, Asselin~de
  Beauville, Marion and Mansot}]{petit2005}
\bibinfo{author}{Petit\xfnm[ R.]}, \bibinfo{author}{Legrand\xfnm[ M.]},
  \bibinfo{author}{Jankowiak\xfnm[ I.]}, \bibinfo{author}{Molini{\'e}\xfnm[
  J.]}, \bibinfo{author}{Asselin~de Beauville\xfnm[ C.]},
  \bibinfo{author}{Marion\xfnm[ G.]}, \bibinfo{author}{Mansot\xfnm[ J.]}.
\newblock \bibinfo{title}{{Transport of Saharan dust over the Caribbean
  Islands: Study of an event}}.
\newblock \bibinfo{journal}{Journal of Geophysical Research: Atmospheres}
  \bibinfo{year}{2005};\bibinfo{volume}{110}(\bibinfo{number}{D18}).
\bibitem[{Pierini et~al.(2012)Pierini, Lovallo and Telesca}]{pierini2012}
\bibinfo{author}{Pierini\xfnm[ J.O.]}, \bibinfo{author}{Lovallo\xfnm[ M.]},
  \bibinfo{author}{Telesca\xfnm[ L.]}.
\newblock \bibinfo{title}{Visibility graph analysis of wind speed records
  measured in central argentina}.
\newblock \bibinfo{journal}{Physica A: Statistical Mechanics and its
  Applications}
  \bibinfo{year}{2012};\bibinfo{volume}{391}(\bibinfo{number}{20}):\bibinfo{pages}{5041--5048}.
\bibitem[{Plocoste and Calif(2019)}]{plocoste2019b}
\bibinfo{author}{Plocoste\xfnm[ T.]}, \bibinfo{author}{Calif\xfnm[ R.]}.
\newblock \bibinfo{title}{{Spectral Observations of PM10 Fluctuations in the
  Hilbert Space}}.
\newblock In: \bibinfo{booktitle}{Functional Calculus}.
  \bibinfo{publisher}{IntechOpen}; \bibinfo{year}{2019}. p.
  \bibinfo{pages}{1--13}.
\bibitem[{Plocoste et~al.(2020{\natexlab{a}})Plocoste, Calif,
  Euphrasie-Clotilde and Brute}]{plocoste2020b}
\bibinfo{author}{Plocoste\xfnm[ T.]}, \bibinfo{author}{Calif\xfnm[ R.]},
  \bibinfo{author}{Euphrasie-Clotilde\xfnm[ L.]}, \bibinfo{author}{Brute\xfnm[
  F.]}.
\newblock \bibinfo{title}{{The statistical behavior of PM10 events over
  guadeloupean archipelago: Stationarity, modelling and extreme events}}.
\newblock \bibinfo{journal}{Atmospheric Research}
  \bibinfo{year}{2020}{\natexlab{a}};\bibinfo{volume}{241}:\bibinfo{pages}{104956}.
\bibitem[{Plocoste et~al.(2020{\natexlab{b}})Plocoste, Calif,
  Euphrasie-Clotilde and Brute}]{plocoste2020c}
\bibinfo{author}{Plocoste\xfnm[ T.]}, \bibinfo{author}{Calif\xfnm[ R.]},
  \bibinfo{author}{Euphrasie-Clotilde\xfnm[ L.]}, \bibinfo{author}{Brute\xfnm[
  F.N.]}.
\newblock \bibinfo{title}{{Investigation of local correlations between
  particulate matter (PM10) and air temperature in the Caribbean basin using
  Ensemble Empirical Mode Decomposition}}.
\newblock \bibinfo{journal}{Atmospheric Pollution Research}
  \bibinfo{year}{2020}{\natexlab{b}};\bibinfo{volume}{11}:\bibinfo{pages}{1692--1704}.
\bibitem[{Plocoste et~al.(2017)Plocoste, Calif and Jacoby-Koaly}]{plocoste2017}
\bibinfo{author}{Plocoste\xfnm[ T.]}, \bibinfo{author}{Calif\xfnm[ R.]},
  \bibinfo{author}{Jacoby-Koaly\xfnm[ S.]}.
\newblock \bibinfo{title}{Temporal multiscaling characteristics of particulate
  matter {$PM10$} and ground-level ozone {$O_3$} concentrations in {Caribbean}
  region}.
\newblock \bibinfo{journal}{Atmospheric Environment}
  \bibinfo{year}{2017};\bibinfo{volume}{169}:\bibinfo{pages}{22--35}.
\bibitem[{Plocoste et~al.(2019)Plocoste, Calif and
  Jacoby-Koaly}]{plocoste2019a}
\bibinfo{author}{Plocoste\xfnm[ T.]}, \bibinfo{author}{Calif\xfnm[ R.]},
  \bibinfo{author}{Jacoby-Koaly\xfnm[ S.]}.
\newblock \bibinfo{title}{Multi-scale time dependent correlation between
  synchronous measurements of ground-level ozone and meteorological parameters
  in the {Caribbean Basin}}.
\newblock \bibinfo{journal}{Atmospheric Environment}
  \bibinfo{year}{2019};\bibinfo{volume}{211}:\bibinfo{pages}{234--246}.
\bibitem[{Plocoste et~al.(2020{\natexlab{c}})Plocoste, Carmona-Cabezas,
  Jim{\'e}nez-Hornero, Guti{\'e}rrez~de Rav{\'e} and Calif}]{plocoste2020d}
\bibinfo{author}{Plocoste\xfnm[ T.]}, \bibinfo{author}{Carmona-Cabezas\xfnm[
  R.]}, \bibinfo{author}{Jim{\'e}nez-Hornero\xfnm[ F.J.]},
  \bibinfo{author}{Guti{\'e}rrez~de Rav{\'e}\xfnm[ E.]},
  \bibinfo{author}{Calif\xfnm[ R.]}.
\newblock \bibinfo{title}{{Multifractal characterisation of particulate matter
  (PM10) time series in the Caribbean basin using visibility graphs}}.
\newblock \bibinfo{journal}{Atmospheric Pollution Research}
  \bibinfo{year}{2020}{\natexlab{c}};\bibinfo{volume}{In press}.
\bibitem[{Plocoste et~al.(2018)Plocoste, Dorville, Monjoly, Jacoby-Koaly and
  Andr{\'e}}]{plocoste2018}
\bibinfo{author}{Plocoste\xfnm[ T.]}, \bibinfo{author}{Dorville\xfnm[ J.F.]},
  \bibinfo{author}{Monjoly\xfnm[ S.]}, \bibinfo{author}{Jacoby-Koaly\xfnm[
  S.]}, \bibinfo{author}{Andr{\'e}\xfnm[ M.]}.
\newblock \bibinfo{title}{{Assessment of Nitrogen Oxides and Ground-Level Ozone
  behavior in a dense air quality station network: Case study in the Lesser
  Antilles Arc}}.
\newblock \bibinfo{journal}{Journal of the Air \& Waste Management Association}
  \bibinfo{year}{2018};\bibinfo{volume}{68}(\bibinfo{number}{12}):\bibinfo{pages}{1278--1300}.
\bibitem[{Plocoste et~al.(2014)Plocoste, Jacoby-Koaly, Molini{\'e} and
  Petit}]{plocoste2014}
\bibinfo{author}{Plocoste\xfnm[ T.]}, \bibinfo{author}{Jacoby-Koaly\xfnm[ S.]},
  \bibinfo{author}{Molini{\'e}\xfnm[ J.]}, \bibinfo{author}{Petit\xfnm[ R.]}.
\newblock \bibinfo{title}{Evidence of the effect of an urban heat island on air
  quality near a landfill}.
\newblock \bibinfo{journal}{Urban Climate}
  \bibinfo{year}{2014};\bibinfo{volume}{10}:\bibinfo{pages}{745--757}.
\bibitem[{Plocoste and
  Pav{\'o}n-Dom{\'\i}nguez(2020{\natexlab{a}})}]{plocoste2020e}
\bibinfo{author}{Plocoste\xfnm[ T.]},
  \bibinfo{author}{Pav{\'o}n-Dom{\'\i}nguez\xfnm[ P.]}.
\newblock \bibinfo{title}{Multifractal detrended cross-correlation analysis of
  wind speed and solar radiation}.
\newblock \bibinfo{journal}{Chaos: An Interdisciplinary Journal of Nonlinear
  Science}
  \bibinfo{year}{2020}{\natexlab{a}};\bibinfo{volume}{30}(\bibinfo{number}{11}):\bibinfo{pages}{113109}.
\bibitem[{Plocoste and
  Pav{\'o}n-Dom{\'\i}nguez(2020{\natexlab{b}})}]{plocoste2020a}
\bibinfo{author}{Plocoste\xfnm[ T.]},
  \bibinfo{author}{Pav{\'o}n-Dom{\'\i}nguez\xfnm[ P.]}.
\newblock \bibinfo{title}{{Temporal scaling study of particulate matter (PM10)
  and solar radiation influences on air temperature in the Caribbean basin
  using a 3D joint multifractal analysis}}.
\newblock \bibinfo{journal}{Atmospheric Environment}
  \bibinfo{year}{2020}{\natexlab{b}};\bibinfo{volume}{222}:\bibinfo{pages}{117115}.
\bibitem[{Posadas et~al.(2001)Posadas, Gim{\'e}nez, Bittelli, Vaz and
  Flury}]{posadas2001}
\bibinfo{author}{Posadas\xfnm[ A.N.]}, \bibinfo{author}{Gim{\'e}nez\xfnm[ D.]},
  \bibinfo{author}{Bittelli\xfnm[ M.]}, \bibinfo{author}{Vaz\xfnm[ C.M.]},
  \bibinfo{author}{Flury\xfnm[ M.]}.
\newblock \bibinfo{title}{Multifractal characterization of soil particle-size
  distributions}.
\newblock \bibinfo{journal}{Soil Science Society of America Journal}
  \bibinfo{year}{2001};\bibinfo{volume}{65}(\bibinfo{number}{5}):\bibinfo{pages}{1361--1367}.
\bibitem[{Prospero and Carlson(1972)}]{prospero1972}
\bibinfo{author}{Prospero\xfnm[ J.M.]}, \bibinfo{author}{Carlson\xfnm[ T.N.]}.
\newblock \bibinfo{title}{{Vertical and areal distribution of Saharan dust over
  the western equatorial North Atlantic Ocean}}.
\newblock \bibinfo{journal}{Journal of Geophysical Research}
  \bibinfo{year}{1972};\bibinfo{volume}{77}(\bibinfo{number}{27}):\bibinfo{pages}{5255--5265}.
\bibitem[{Prospero and Carlson(1981)}]{prospero1981}
\bibinfo{author}{Prospero\xfnm[ J.M.]}, \bibinfo{author}{Carlson\xfnm[ T.N.]}.
\newblock \bibinfo{title}{{Saharan air outbreaks over the tropical North
  Atlantic}}.
\newblock In: \bibinfo{booktitle}{Weather and Weather Maps}.
  \bibinfo{publisher}{Springer}; \bibinfo{year}{1981}. p.
  \bibinfo{pages}{677--691}.
\bibitem[{Prospero et~al.(2014)Prospero, Collard, Molini{\'e} and
  Jeannot}]{prospero2014}
\bibinfo{author}{Prospero\xfnm[ J.M.]}, \bibinfo{author}{Collard\xfnm[ F.X.]},
  \bibinfo{author}{Molini{\'e}\xfnm[ J.]}, \bibinfo{author}{Jeannot\xfnm[ A.]}.
\newblock \bibinfo{title}{Characterizing the annual cycle of {African} dust
  transport to the {Caribbean Basin} and {South America} and its impact on the
  environment and air quality}.
\newblock \bibinfo{journal}{Global Biogeochemical Cycles}
  \bibinfo{year}{2014};\bibinfo{volume}{28}:\bibinfo{pages}{757--773}.
\bibitem[{Prospero and Lamb(2003)}]{prospero2003}
\bibinfo{author}{Prospero\xfnm[ J.M.]}, \bibinfo{author}{Lamb\xfnm[ P.J.]}.
\newblock \bibinfo{title}{{African droughts and dust transport to the
  Caribbean: Climate change implications}}.
\newblock \bibinfo{journal}{Science}
  \bibinfo{year}{2003};\bibinfo{volume}{302}(\bibinfo{number}{5647}):\bibinfo{pages}{1024--1027}.
\bibitem[{Rastelli et~al.(2017)Rastelli, Corinaldesi, Dell’Anno, Martire,
  Greco, Facchini, Rinaldi, O’Dowd, Ceburnis and Danovaro}]{rastelli2017}
\bibinfo{author}{Rastelli\xfnm[ E.]}, \bibinfo{author}{Corinaldesi\xfnm[ C.]},
  \bibinfo{author}{Dell’Anno\xfnm[ A.]}, \bibinfo{author}{Martire\xfnm[
  M.L.]}, \bibinfo{author}{Greco\xfnm[ S.]}, \bibinfo{author}{Facchini\xfnm[
  M.C.]}, \bibinfo{author}{Rinaldi\xfnm[ M.]}, \bibinfo{author}{O’Dowd\xfnm[
  C.]}, \bibinfo{author}{Ceburnis\xfnm[ D.]}, \bibinfo{author}{Danovaro\xfnm[
  R.]}.
\newblock \bibinfo{title}{Transfer of labile organic matter and microbes from
  the ocean surface to the marine aerosol: an experimental approach}.
\newblock \bibinfo{journal}{Scientific reports}
  \bibinfo{year}{2017};\bibinfo{volume}{7}(\bibinfo{number}{1}):\bibinfo{pages}{11475}.
\bibitem[{Sannino et~al.(2017)Sannino, Stramaglia, Lacasa and
  Marinazzo}]{sannino2017}
\bibinfo{author}{Sannino\xfnm[ S.]}, \bibinfo{author}{Stramaglia\xfnm[ S.]},
  \bibinfo{author}{Lacasa\xfnm[ L.]}, \bibinfo{author}{Marinazzo\xfnm[ D.]}.
\newblock \bibinfo{title}{{Visibility graphs for fMRI data: Multiplex temporal
  graphs and their modulations across resting-state networks}}.
\newblock \bibinfo{journal}{Network Neuroscience}
  \bibinfo{year}{2017};\bibinfo{volume}{1}(\bibinfo{number}{3}):\bibinfo{pages}{208--221}.
\bibitem[{Scheers et~al.(2015)Scheers, Jacobs, Casas, Nemery and
  Nawrot}]{scheers2015}
\bibinfo{author}{Scheers\xfnm[ H.]}, \bibinfo{author}{Jacobs\xfnm[ L.]},
  \bibinfo{author}{Casas\xfnm[ L.]}, \bibinfo{author}{Nemery\xfnm[ B.]},
  \bibinfo{author}{Nawrot\xfnm[ T.S.]}.
\newblock \bibinfo{title}{{Long-term exposure to particulate matter air
  pollution is a risk factor for stroke: Meta-analytical evidence}}.
\newblock \bibinfo{journal}{Stroke}
  \bibinfo{year}{2015};\bibinfo{volume}{46}(\bibinfo{number}{11}):\bibinfo{pages}{3058--3066}.
\bibitem[{Schepanski(2018)}]{schepanski2018}
\bibinfo{author}{Schepanski\xfnm[ K.]}.
\newblock \bibinfo{title}{Transport of mineral dust and its impact on climate}.
\newblock \bibinfo{journal}{Geosciences}
  \bibinfo{year}{2018};\bibinfo{volume}{8}(\bibinfo{number}{5}):\bibinfo{pages}{151}.
\bibitem[{Schmitt(2005)}]{schmitt2005}
\bibinfo{author}{Schmitt\xfnm[ F.G.]}.
\newblock \bibinfo{title}{Explicit predictability and dispersion scaling
  exponents in fully developed turbulence}.
\newblock \bibinfo{journal}{Physics Letters A}
  \bibinfo{year}{2005};\bibinfo{volume}{342}(\bibinfo{number}{5-6}):\bibinfo{pages}{448--458}.
\bibitem[{Schreiber and Grussbach(1991)}]{schreiber1991}
\bibinfo{author}{Schreiber\xfnm[ M.]}, \bibinfo{author}{Grussbach\xfnm[ H.]}.
\newblock \bibinfo{title}{{Multifractal wave functions at the Anderson
  transition}}.
\newblock \bibinfo{journal}{Physical review letters}
  \bibinfo{year}{1991};\bibinfo{volume}{67}(\bibinfo{number}{5}):\bibinfo{pages}{607}.
\bibitem[{Schwartz(1995)}]{schwartz1995}
\bibinfo{author}{Schwartz\xfnm[ J.]}.
\newblock \bibinfo{title}{Short term fluctuations in air pollution and hospital
  admissions of the elderly for respiratory disease.}
\newblock \bibinfo{journal}{Thorax}
  \bibinfo{year}{1995};\bibinfo{volume}{50}(\bibinfo{number}{5}):\bibinfo{pages}{531--538}.
\bibitem[{Seuront et~al.(1996)Seuront, Schmitt, Lagadeuc, Schertzer, Lovejoy
  and Frontier}]{seuront1996}
\bibinfo{author}{Seuront\xfnm[ L.]}, \bibinfo{author}{Schmitt\xfnm[ F.]},
  \bibinfo{author}{Lagadeuc\xfnm[ Y.]}, \bibinfo{author}{Schertzer\xfnm[ D.]},
  \bibinfo{author}{Lovejoy\xfnm[ S.]}, \bibinfo{author}{Frontier\xfnm[ S.]}.
\newblock \bibinfo{title}{Multifractal analysis of phytoplankton biomass and
  temperature in the ocean}.
\newblock \bibinfo{journal}{Geophysical Research Letters}
  \bibinfo{year}{1996};\bibinfo{volume}{23}(\bibinfo{number}{24}):\bibinfo{pages}{3591--3594}.
\bibitem[{Soni(2019)}]{soni2019}
\bibinfo{author}{Soni\xfnm[ G.]}.
\newblock \bibinfo{title}{Signed visibility graphs of time series and their
  application to brain networks}.
\newblock Ph.D. thesis; University of British Columbia; \bibinfo{year}{2019}.
\bibitem[{T{\'e}l et~al.(1989)T{\'e}l, F{\"u}l{\"o}p and Vicsek}]{tel1989}
\bibinfo{author}{T{\'e}l\xfnm[ T.]}, \bibinfo{author}{F{\"u}l{\"o}p\xfnm[
  {\'A}.]}, \bibinfo{author}{Vicsek\xfnm[ T.]}.
\newblock \bibinfo{title}{Determination of fractal dimensions for geometrical
  multifractals}.
\newblock \bibinfo{journal}{Physica A: Statistical Mechanics and its
  Applications}
  \bibinfo{year}{1989};\bibinfo{volume}{159}(\bibinfo{number}{2}):\bibinfo{pages}{155--166}.
\bibitem[{Tessier et~al.(1994)Tessier, Lovejoy and Schertzer}]{tessier1994}
\bibinfo{author}{Tessier\xfnm[ Y.]}, \bibinfo{author}{Lovejoy\xfnm[ S.]},
  \bibinfo{author}{Schertzer\xfnm[ D.]}.
\newblock \bibinfo{title}{Multifractal analysis and simulation of the global
  meteorological network}.
\newblock \bibinfo{journal}{Journal of applied meteorology}
  \bibinfo{year}{1994};\bibinfo{volume}{33}(\bibinfo{number}{12}):\bibinfo{pages}{1572--1586}.
\bibitem[{Turner et~al.(2001)Turner, Doxa, O'sullivan and Penn}]{turner2001}
\bibinfo{author}{Turner\xfnm[ A.]}, \bibinfo{author}{Doxa\xfnm[ M.]},
  \bibinfo{author}{O'sullivan\xfnm[ D.]}, \bibinfo{author}{Penn\xfnm[ A.]}.
\newblock \bibinfo{title}{From isovists to visibility graphs: a methodology for
  the analysis of architectural space}.
\newblock \bibinfo{journal}{Environment and Planning B: Planning and design}
  \bibinfo{year}{2001};\bibinfo{volume}{28}(\bibinfo{number}{1}):\bibinfo{pages}{103--121}.
\bibitem[{Van Der~Does et~al.(2016)Van Der~Does, Korte, Munday, Brummer and
  Stuut}]{van2016}
\bibinfo{author}{Van Der~Does\xfnm[ M.]}, \bibinfo{author}{Korte\xfnm[ L.F.]},
  \bibinfo{author}{Munday\xfnm[ C.I.]}, \bibinfo{author}{Brummer\xfnm[
  G.J.A.]}, \bibinfo{author}{Stuut\xfnm[ J.B.W.]}.
\newblock \bibinfo{title}{{Particle size traces modern Saharan dust transport
  and deposition across the equatorial North Atlantic}}.
\newblock \bibinfo{journal}{Atmospheric Chemistry \& Physics}
  \bibinfo{year}{2016};\bibinfo{volume}{16}(\bibinfo{number}{21}).
\bibitem[{Velasco-Merino et~al.(2018)Velasco-Merino, Mateos, Toledano,
  Prospero, Molinie, Euphrasie-Clotilde, Gonz{\'a}lez, Cachorro, Calle and
  Frutos}]{velasco2018}
\bibinfo{author}{Velasco-Merino\xfnm[ C.]}, \bibinfo{author}{Mateos\xfnm[ D.]},
  \bibinfo{author}{Toledano\xfnm[ C.]}, \bibinfo{author}{Prospero\xfnm[ J.M.]},
  \bibinfo{author}{Molinie\xfnm[ J.]},
  \bibinfo{author}{Euphrasie-Clotilde\xfnm[ L.]},
  \bibinfo{author}{Gonz{\'a}lez\xfnm[ R.]}, \bibinfo{author}{Cachorro\xfnm[
  V.E.]}, \bibinfo{author}{Calle\xfnm[ A.]}, \bibinfo{author}{Frutos\xfnm[
  A.M.d.]}.
\newblock \bibinfo{title}{Impact of long-range transport over the {Atlantic
  Ocean on Saharan dust optical and microphysical properties based on AERONET
  data}}.
\newblock \bibinfo{journal}{Atmospheric Chemistry and Physics}
  \bibinfo{year}{2018};\bibinfo{volume}{18}(\bibinfo{number}{13}):\bibinfo{pages}{9411--9424}.
\bibitem[{Veneziano et~al.(1995)Veneziano, Moglen and Bras}]{veneziano1995}
\bibinfo{author}{Veneziano\xfnm[ D.]}, \bibinfo{author}{Moglen\xfnm[ G.E.]},
  \bibinfo{author}{Bras\xfnm[ R.L.]}.
\newblock \bibinfo{title}{Multifractal analysis: pitfalls of standard
  procedures and alternatives}.
\newblock \bibinfo{journal}{Physical Review E}
  \bibinfo{year}{1995};\bibinfo{volume}{52}(\bibinfo{number}{2}):\bibinfo{pages}{1387}.
\bibitem[{Vicsek et~al.(1990)Vicsek, Family and Meakin}]{vicsek1990}
\bibinfo{author}{Vicsek\xfnm[ T.]}, \bibinfo{author}{Family\xfnm[ F.]},
  \bibinfo{author}{Meakin\xfnm[ P.]}.
\newblock \bibinfo{title}{Multifractal geometry of diffusion-limited
  aggregates}.
\newblock \bibinfo{journal}{EPL (Europhysics Letters)}
  \bibinfo{year}{1990};\bibinfo{volume}{12}(\bibinfo{number}{3}):\bibinfo{pages}{217}.
\bibitem[{Viel et~al.(2020)Viel, Michineau, Garbin, Monfort, Kadhel, Multigner
  and Rouget}]{viel2020}
\bibinfo{author}{Viel\xfnm[ J.F.]}, \bibinfo{author}{Michineau\xfnm[ L.]},
  \bibinfo{author}{Garbin\xfnm[ C.]}, \bibinfo{author}{Monfort\xfnm[ C.]},
  \bibinfo{author}{Kadhel\xfnm[ P.]}, \bibinfo{author}{Multigner\xfnm[ L.]},
  \bibinfo{author}{Rouget\xfnm[ F.]}.
\newblock \bibinfo{title}{Impact of saharan dust on severe small for
  gestational births in the caribbean}.
\newblock \bibinfo{journal}{The American Journal of Tropical Medicine and
  Hygiene}
  \bibinfo{year}{2020};\bibinfo{volume}{102}(\bibinfo{number}{6}):\bibinfo{pages}{1463--1465}.
\bibitem[{Woodruff et~al.(1997)Woodruff, Grillo and Schoendorf}]{woodruff1997}
\bibinfo{author}{Woodruff\xfnm[ T.J.]}, \bibinfo{author}{Grillo\xfnm[ J.]},
  \bibinfo{author}{Schoendorf\xfnm[ K.C.]}.
\newblock \bibinfo{title}{{The relationship between selected causes of
  postneonatal infant mortality and particulate air pollution in the United
  States}}.
\newblock \bibinfo{journal}{Environmental health perspectives}
  \bibinfo{year}{1997};\bibinfo{volume}{105}(\bibinfo{number}{6}):\bibinfo{pages}{608--612}.
\bibitem[{Yang(2002)}]{yang2002}
\bibinfo{author}{Yang\xfnm[ K.L.]}.
\newblock \bibinfo{title}{{Spatial and seasonal variation of PM10 mass
  concentrations in Taiwan}}.
\newblock \bibinfo{journal}{Atmospheric Environment}
  \bibinfo{year}{2002};\bibinfo{volume}{36}(\bibinfo{number}{21}):\bibinfo{pages}{3403--3411}.
\bibitem[{Yu et~al.(2016)Yu, Zhang, Huang, Lin and Anh}]{yu2016}
\bibinfo{author}{Yu\xfnm[ Z.G.]}, \bibinfo{author}{Zhang\xfnm[ H.]},
  \bibinfo{author}{Huang\xfnm[ D.W.]}, \bibinfo{author}{Lin\xfnm[ Y.]},
  \bibinfo{author}{Anh\xfnm[ V.]}.
\newblock \bibinfo{title}{{Multifractality and Laplace spectrum of horizontal
  visibility graphs constructed from fractional Brownian motions}}.
\newblock \bibinfo{journal}{Journal of Statistical Mechanics: Theory and
  Experiment}
  \bibinfo{year}{2016};\bibinfo{volume}{2016}(\bibinfo{number}{3}):\bibinfo{pages}{033206}.
\bibitem[{Zhang et~al.(2017)Zhang, Liu, Cui, Liu, Yin and Li}]{zhang2017}
\bibinfo{author}{Zhang\xfnm[ J.]}, \bibinfo{author}{Liu\xfnm[ Y.]},
  \bibinfo{author}{Cui\xfnm[ L.l.]}, \bibinfo{author}{Liu\xfnm[ S.q.]},
  \bibinfo{author}{Yin\xfnm[ X.x.]}, \bibinfo{author}{Li\xfnm[ H.c.]}.
\newblock \bibinfo{title}{{Ambient air pollution, smog episodes and mortality
  in Jinan, China}}.
\newblock \bibinfo{journal}{Scientific reports}
  \bibinfo{year}{2017};\bibinfo{volume}{7}(\bibinfo{number}{1}):\bibinfo{pages}{1--8}.

\end{thebibliography}

\end{document}